\documentclass[aps,prl,twocolumn,superscriptaddress]{revtex4}

\usepackage[dvips]{graphicx}
\usepackage{color}
\usepackage{bm}
\usepackage{amssymb}

\usepackage[utf8]{inputenc}
\usepackage{times}

\begin{document}

\title{Spin-orbit coupling controlled ground states in the double perovskite iridates $A_2 B$IrO$_6$ ($A = $ Ba, Sr; $B = $~Lu, Sc)}

\author{A.A. Aczel}
\email{aczelaa@ornl.gov}
\affiliation{Neutron Scattering Division, Oak Ridge National Laboratory, Oak Ridge, TN 37831, USA}

\author{Q. Chen}
\affiliation{Department of Physics and Astronomy, University of Tennessee, Knoxville, TN 37996, USA}

\author{J.P. Clancy}
\affiliation{Department of Physics and Astronomy, McMaster University, Hamilton, ON L8S 4M1, Canada}
\affiliation{Brockhouse Institute for Materials Research, McMaster University, Hamilton, ON L8S 4M1, Canada}

\author{C. dela Cruz}
\affiliation{Neutron Scattering Division, Oak Ridge National Laboratory, Oak Ridge, TN 37831, USA}

\author{D. Reig-i-Plessis}
\affiliation{Department of Physics, University of Illinois at Urbana-Champaign, Urbana, IL 61801, USA}

\author{G.J. MacDougall}
\affiliation{Department of Physics, University of Illinois at Urbana-Champaign, Urbana, IL 61801, USA}

\author{C.J. Pollock}
\affiliation{Cornell High Energy Synchrotron Source, Cornell University, Ithaca, NY 14853, USA}

\author{M.H. Upton}
\affiliation{Advanced Photon Source, Argonne National Laboratory, Lemont, IL 60439, USA}

\author{T.J. Williams}
\affiliation{Neutron Scattering Division, Oak Ridge National Laboratory, Oak Ridge, TN 37831, USA}

\author{N. LaManna}
\affiliation{Department of Physics, Villanova University, Villanova, PA 19085, USA}

\author{J.P. Carlo}
\affiliation{Department of Physics, Villanova University, Villanova, PA 19085, USA}

\author{J. Beare}
\affiliation{Department of Physics and Astronomy, McMaster University, Hamilton, ON L8S 4M1, Canada}

\author{G.M. Luke}
\affiliation{Department of Physics and Astronomy, McMaster University, Hamilton, ON L8S 4M1, Canada}

\author{H.D. Zhou}
\affiliation{Department of Physics and Astronomy, University of Tennessee, Knoxville, TN 37996, USA}

\date{\today}

\begin{abstract}
Iridates with the 5$d^4$ electronic configuration have attracted recent interest due to reports of magnetically-ordered ground states despite longstanding expectations that their strong spin-orbit coupling would generate a $J = 0$ electronic ground state for each Ir$^{5+}$ ion. The major focus of prior research has been on the double perovskite iridates Ba$_2$YIrO$_6$ and Sr$_2$YIrO$_6$, where the nature of the ground states (i.e. ordered vs non-magnetic) is still controversial. Here we present neutron powder diffraction, high energy resolution fluorescence detected x-ray absorption spectroscopy (HERFD-XAS), resonant inelastic x-ray scattering (RIXS), magnetic susceptibility, and muon spin relaxation data on the related double perovskite iridates Ba$_2$LuIrO$_6$, Sr$_2$LuIrO$_6$, Ba$_2$ScIrO$_6$, and Sr$_2$ScIrO$_6$ that enable us to gain a general understanding of the electronic and magnetic properties for this family of materials. Our HERFD-XAS and RIXS measurements establish $J = 0$ electronic ground states for the Ir$^{5+}$ ions in all cases, with similar values for Hund's coupling $J_{\rm H}$ and the spin-orbit coupling constant $\lambda_{\rm SOC}$. Our bulk susceptibility and muon spin relaxation data find no evidence for long-range magnetic order or spin freezing, but they do exhibit weak magnetic signals that are consistent with extrinsic local moments. Our results indicate that the large $\lambda_{\rm SOC}$ is the key driving force behind the electronic and magnetic ground states realized in the 5$d^4$ double perovskite iridates, which agrees well with conventional wisdom. 
\end{abstract}

\maketitle

\section{I. Introduction}

Heavy transition metal magnets based on 4$d$ and 5$d$ ions offer opportunities to identify and characterize exotic magnetic states and phenomena due to comparable and competing energy scales for the spin-orbit coupling (SOC), electronic correlations, and crystal field effects \cite{14_witczak, 16_du, 16_rau, 16_schaffer, 17_martins, 18_cao, 21_takayama}. An octahedral local environment for the magnetic ions leads to particularly rich physics. For example, iridates with the 5$d^5$ electronic configuration have been of enduring interest due to their penchant for hosting $J_{\rm eff} =$~1/2 spin-orbit-assisted Mott insulating ground states \cite{08_kim} that can give rise to Kitaev spin liquid physics \cite{16_banerjee, 17_banerjee, 17_winter, 18_hermanns, 19_takagi} and offer promise in the ongoing search for new superconductors \cite{11_wang, 13_watanabe}. The phase diagrams for materials with 4$d^1$/5$d^1$ and 4$d^2$/5$d^2$ magnetic ions are also quite complex \cite{10_chen, 11_chen, 11_dodds, 14_ishizuka, 21_svoboda}, with multipolar order and valence bond glass states both predicted \cite{17_romhanyi, 20_paramekanti, 20_voleti, 21_khaliullin, 21_mosca, 21_pourosvskii} and realized in the laboratory \cite{10_aharen, 10_devries, 20_hirai, 20_maharaj, 21_lee, 21_mustonen}. 

While systems with transition metals in the 5$d^4$ electronic configuration are expected to host $J = 0$ electronic ground states due to completely filled $j_{\rm eff} = 3/2$ manifolds arising from the large SOC, this assumption has been challenged recently by both experiment \cite{14_cao, 17_terzic, 20_laguna} and theory \cite{13_khaliullin, 15_bhowal, 15_meetei}. Features typically associated with magnetic order have been identified in bulk characterization measurements on Ba$_2$YIrO$_6$ \cite{17_terzic} and Sr$_2$YIrO$_6$ \cite{14_cao}, including a sharp peak in the temperature dependence of the magnetic susceptibility, an upturn in the field dependence of the magnetization, and a broad peak in the specific heat. Oscillations have also been observed in zero-field muon spin relaxation ($\mu$SR) data on Sr$_2$YIrO$_6$ \cite{20_laguna} and correspond to 17\% of the volume fraction. Finally, density functional theory (DFT) calculations predicted A-type antiferromagnetic ground states for both Ba$_2$YIrO$_6$ and Sr$_2$YIrO$_6$ \cite{15_bhowal}. 

Two mechanisms have been proposed for generating magnetic order in 5$d^4$ systems. The first possibility is realized in the limit where the non-cubic crystal field $\Delta >> \lambda_{\rm SOC}$, as an octahedral compression leads to quenching of the orbital angular momentum and generates an $S =$~1 Heisenberg magnet \cite{14_cao}. Previous work has shown that this scenario applies to some 4$d^4$ magnets, including Ca$_2$RuO$_4$ \cite{17_jain}. However, it cannot account for magnetic order in cubic Ba$_2$YIrO$_6$ with ideal IrO$_6$ octahedra. The second possibility applies more generally to all 5$d^4$ systems and arises from competition between $\lambda_{\rm SOC}$ and superexchange interactions. In the atomic limit, the expected single-ion ground state is a $J =$~0 singlet with a $J_{\rm eff} =$~1 triplet excited state. The introduction of superexchange allows the $J_{\rm eff} =$~1 excitations to hop between sites and hence acquire dispersion. As superexchange increases, the energy gap between the non-magnetic $J =$~0 ground state and the $J_{\rm eff} =$~1 excitation band decreases until a phase transition to a magnetic state is realized. This phenomenon is known as Van Vleck-type singlet-triplet excitonic magnetism or ``excitonic" condensation \cite{13_khaliullin, 19_kim}. 

Confirmation of magnetic order in Ba$_2$YIrO$_6$  and Sr$_2$YIrO$_6$ has remained elusive, as it has not been possible to reproduce the key low-temperature features observed in the magnetic susceptibility and magnetization measurements, and the $\mu$SR result for Sr$_2$YIrO$_6$ is inconsistent with the susceptibility data presented in the same paper \cite{20_laguna}. Most previous studies have concluded that these materials host non-magnetic ground states instead \cite{16_dey, 16_pajskr, 16_phelan, 17_chen, 17_corredor, 17_hammerath, 18_fuchs, 18_gong, 18_kusch}, with sample purity issues possibly contributing to the anomalous results. Although many probes find evidence for weak paramagnetic signals, they are typically attributed to extrinsic sources. Proposed origins include paramagnetic impurities \cite{17_corredor, 17_hammerath}, the presence of Ir$^{4+}$ or Ir$^{6+}$ ions due to off-stoichiometry \cite{16_dey, 18_fuchs}, and antisite disorder \cite{17_chen}. In fact, it was shown that the broad peak observed in the specific heat of Sr$_2$YIrO$_6$ could be attributed to a Schottky anomaly associated with a small paramagnetic impurity fraction \cite{17_corredor}. A subsequent DFT study on Ba$_2$YIrO$_6$ and Sr$_2$YIrO$_6$ also found lower-energy non-magnetic states that were not considered in Ref.~\cite{15_bhowal}. Finally, resonant inelastic x-ray scattering and theoretical calculations on Ba$_2$YIrO$_6$ find a singlet-triplet gap of 350 - 370 meV and a weak triplet dispersion $<$~50~meV \cite{17_chen, 18_kusch}, which effectively rules out an excitonic condensation mechanism. 

With mounting evidence for non-magnetic ground states in Ba$_2$YIrO$_6$ and Sr$_2$YIrO$_6$, it is unclear if magnetic order can be induced in 5$d^4$ systems via either of the two mechanisms described above. Motivated by this open question, here we investigate the crystal structures, single-ion properties, and collective magnetic ground states of the 5$d^4$ double perovskite iridates Ba$_2$LuIrO$_6$, Ba$_2$ScIrO$_6$, Sr$_2$LuIrO$_6$, and Sr$_2$ScIrO$_6$ with a combination of neutron powder diffraction (NPD), high energy resolution fluorescence detected x-ray absorption spectroscopy (HERFD-XAS), resonant inelastic x-ray scattering (RIXS), magnetic susceptibility, and muon spin relaxation ($\mu$SR) measurements. The diffraction results indicate that the Ba and Sr systems crystallize in the cubic {\it Fm$\bar{3}$m} and monoclinic {\it P$2_1$/n} space groups respectively. The HERFD-XAS measurements show that the oxidation state in these four compounds is predominantly Ir$^{5+}$, suggesting that any Ir$^{4+}$ or Ir$^{6+}$ defects must occur at small concentrations. The RIXS measurements reveal similar values of $\lambda_{\rm SOC}$ and Hund's coupling $J_{\rm H}$ for all four systems that are comparable to previous estimates for Ba$_2$YIrO$_6$ \cite{18_nag} and Sr$_2$YIrO$_6$ \cite{17_yuan}. Most importantly, our combined susceptibility and $\mu$SR results find no evidence for long-range magnetic order or spin freezing in any of these compounds down to 25~mK. We attribute the weak magnetic signals in these data to local moments arising from a small amount of a 6L-perovskite polytype impurity phase or antisite disorder found in the diffraction measurements on the Ba and Sr samples respectively, as well as small concentrations of Ir$^{4+}$ and/or Ir$^{6+}$ defects identified in the RIXS spectra of all four samples. Our results indicate that $\lambda_{\rm SOC}$ is the dominant energy scale in these materials and drives the formation of a non-magnetic $J =$~0 electronic ground state at the Ir$^{5+}$ sites. 

\section{II. Experimental details}

Polycrystalline samples of Ba$_2$YIrO$_6$, Ba$_2$LuIrO$_6$, Ba$_2$ScIrO$_6$, Sr$_2$YIrO$_6$, Sr$_2$LuIrO$_6$, and Sr$_2$ScIrO$_6$ were synthesized by conventional solid state reactions. For all target phases, appropriate stoichiometric amounts of the starting materials BaCO$_3$, SrCO$_3$, Ir (with 5 wt\% extra added), Y$_2$O$_3$, Lu$_2$O$_3$, and Sc$_2$O$_3$ with purities not less than 99.9\% were first mixed together. The homogeneous powders were then pelletized, placed in a covered alumina crucible, and heated at 900$^\circ$C for 6 hours. The pellets were then ground, re-pelletized, and fired at 1050$^\circ$C and then 1250$^\circ$C for 20 hours with intermediate grinding. Finally, one last sintering was performed at 1250$^\circ$C for 40 hours to obtain Sr$_2$YIrO$_6$, 1300$^\circ$C for 15 hours to obtain Ba$_2$YIrO$_6$, Ba$_2$LuIrO$_6$ and Sr$_2$LuIrO$_6$, 1425$^\circ$C for 15 hours to obtain Sr$_2$ScIrO$_6$, and 1450$^\circ$C for 15 hours to obtain Ba$_2$ScIrO$_6$. Phase verification was completed by collecting room temperature X-ray diffraction (XRD) patterns with a HUBER Imaging Plate Guinier Camera 670 using Cu $K_{\alpha1}$ radiation (1.54~\AA). Magnetic susceptibility measurements were obtained with a vibrating sample magnetometer. 

Neutron powder diffraction (NPD) was performed with $\sim$~2 g of polycrystalline Ba$_2$LuIrO$_6$, Ba$_2$ScIrO$_6$, Sr$_2$LuIrO$_6$, and Sr$_2$ScIrO$_6$, as well as a reference sample of Ba$_2$YIrO$_6$, using the HB-2A powder diffractometer \cite{18_calder} of the High Flux Isotope Reactor (HFIR) at Oak Ridge National Laboratory (ORNL) to revisit the crystal structures of these materials systematically. The samples were loaded in cylindrical vanadium cans with 5~mm inner diameters. The data were collected at $T$~$=$~4~K with a neutron wavelength of 1.54~\AA~and a collimation of open-21$'$-12$'$. 

High energy resolution fluorescence detected x-ray absorption spectroscopy (HERFD-XAS) measurements were performed on polycrystalline samples of Ba$_2$LuIrO$_6$, Ba$_2$ScIrO$_6$, Sr$_2$LuIrO$_6$, and Sr$_2$ScIrO$_6$, as well as reference samples of IrO$_2$ and Ba$_2$CeIrO$_6$ (nominally Ir$^{4+}$) and Ba$_2$YIrO$_6$ and Sr$_2$YIrO$_6$ (nominally Ir$^{5+}$). Measurements were carried out at room temperature using the PIPOXS (ID2A) beamline at the Cornell High Energy Synchrotron Source (CHESS). These measurements employed a liquid nitrogen-cooled Si-(3 1 1) double crystal monochromator, and five spherical (1 m radius) Si-(8 0 0) analyzers arranged in Rowland geometry. By tuning the incident x-ray energy to the Ir $L_3$ absorption edge ($E_i$ = 11.215 keV) XAS can be used to probe the valence state of the Ir ions in these compounds. By limiting the emitted photon energies to a narrow range, HERFD-XAS can circumvent Ir core-hole lifetime broadening, and provide significantly higher energy resolution ($\sim$2.7 eV) than conventional XAS at the Ir $L_3$ edge ($\sim$5.3 eV).

Resonant inelastic x-ray scattering (RIXS) measurements were conducted on polycrystalline samples of Ba$_2$LuIrO$_6$, Ba$_2$ScIrO$_6$, Sr$_2$LuIrO$_6$, and Sr$_2$ScIrO$_6$, as well as reference samples of Ba$_2$YIrO$_6$ and Sr$_2$YIrO$_6$, using the MERIX spectrometer on beamline 27-ID of the Advanced Photon Source (APS) at Argonne National Laboratory to investigate the Ir$^{5+}$ excitation spectra. The incident x-ray energy was tuned to the Ir $L_3$ absorption edge at 11.215 keV. A double-bounce diamond-(1 1 1) primary monochromator, a channel-cut Si-(8 4 4) secondary monochromator, and a spherical (2 m radius) diced Si-(8 4 4) analyzer crystal were used to obtain an overall energy resolution of $\sim$~40 meV (full width at half maximum [FWHM]). In order to minimize the elastic background intensity, measurements were carried out in horizontal scattering geometry with the scattering angle 2$\theta$ set to 90$^\circ$. All measurements were carried out at room temperature.

Muon spin relaxation ($\mu$SR) measurements were performed at TRIUMF, Canada on polycrystalline samples of Ba$_2$LuIrO$_6$, Ba$_2$ScIrO$_6$, Sr$_2$LuIrO$_6$, and Sr$_2$ScIrO$_6$, as well as a reference sample of Ba$_2$YIrO$_6$, using the M15 surface muon beamline equipped with the Dilution Refrigerator (DR), which has a base temperature of 25~mK. To improve thermal contact with the sample holder, the samples were mixed with 14 - 18 wt\% Ag powder and then pressed into 1~g pellets. In a $\mu$SR experiment, positive muons with a mean lifetime of 2.2  $\mu s$ stop in the sample one at a time with the location determined by the minimum in the Coulomb energy. The muon spins then precess around the local magnetic field at the stopping site before decaying into a positron plus two neutrinos, with the positron emitted preferentially along the muon spin direction at the time of decay. For a fixed set of experimental conditions (i.e. temperature and magnetic field), millions of positron events are recorded by two opposing counters, which allows one to determine the time-evolution of the muon spin polarization. This quantity is known as the muon asymmetry \cite{97_dereotier, yaouanc_textbook} and is given by: 
\begin{equation}
A(t) = \frac{N_B(t) - \alpha N_F(t)}{N_B(t) + \alpha N_F(t)} , 
\end{equation}
where $N_B(t)$ and $N_F(t)$ are the number of counts in the two (back and front) positron counters and $\alpha$ is a calibration parameter which was determined via a weak transverse field measurement at 7 K. Our other measurements were performed in zero-field (ZF) and longitudinal-field (LF) geometry. 

\section{III. Crystal structures}

$B/B'$-site ordered double perovskites of the form $A_2 B B'$O$_6$ can crystallize in the ideal {\it Fm$\bar{3}$m} crystal structure or lower-symmetry variants. The ionic radius $r$ of the $A$-site typically plays a key role in crystal structure determination, with many Ba ($r =$~1.42~\AA) systems forming the ideal face-centered-cubic structure and Sr ($r =$~1.26~\AA) systems often crystallizing in the monoclinic {\it P$2_1$/n} space group. Several diffraction studies have now shown that this general trend seems to apply to Ba$_2$YIrO$_6$ \cite{15_ranjbar, 16_dey, 17_chen, 17_terzic, 18_nag} and Sr$_2$YIrO$_6$ \cite{99_wakeshima, 14_cao, 15_ranjbar, 20_laguna}, although there are reports of Ba$_2$YIrO$_6$ crystallizing in the space group {\it P$2_1$/n} \cite{99_wakeshima} and Sr$_2$YIrO$_6$ crystallizing in the space group {\it Fm$\bar{3}$m} \cite{17_corredor}. There has been less focus on the related 5$d^4$ double perovskite iridates Ba$_2$LuIrO$_6$, Sr$_2$LuIrO$_6$, Ba$_2$ScIrO$_6$, and Sr$_2$ScIrO$_6$, although recent synchrotron x-ray diffraction work on the Sc samples also identifies cubic and monoclinic crystal symmetry for the Ba and Sr analogs \cite{17_kayser}. The only prior diffraction data available for Ba$_2$LuIrO$_6$ and Sr$_2$LuIrO$_6$ came from a laboratory x-ray source, which is rather insensitive to oxygen, and the crystal symmetry of both systems is reported to be monoclinic \cite{99_wakeshima}. 

\begin{figure}
\centering
\scalebox{0.17}{\includegraphics{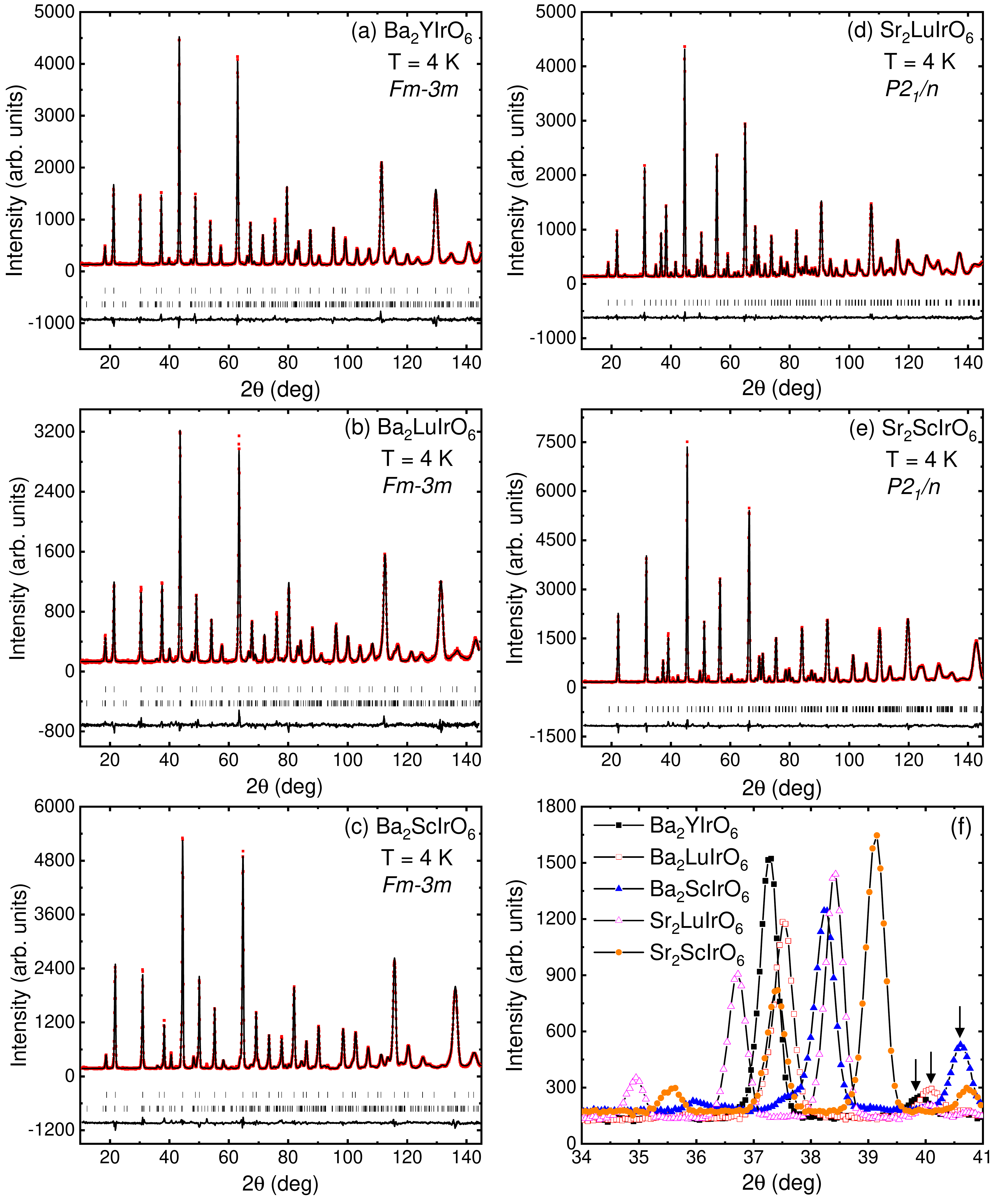}}
\caption{\label{Fig1} (color online) Neutron powder diffraction data, indicated by the solid red squares and collected with a neutron wavelength~1.54~\AA~at a temperature $T$~$=$~4~K, is shown for (a) Ba$_2$YIrO$_6$, (b) Ba$_2$LuIrO$_6$, (c) Ba$_2$ScIrO$_6$, (d) Sr$_2$LuIrO$_6$, and (e) Sr$_2$ScIrO$_6$. The Rietveld refinement results are superimposed on the data as solid black curves, the difference curves are shown below the diffraction patterns, and the expected Bragg peak positions are indicated by ticks. (f) A zoomed-in version of a small portion of the diffraction patterns for all five materials. There are fewer prominent Bragg peaks observed in the data for the Ba samples as compared to the data for the Sr samples, which is indicative of a higher crystal symmetry for Ba$_2$YIrO$_6$, Ba$_2$LuIrO$_6$, and Ba$_2$ScIrO$_6$. The Bragg peaks marked by the black arrows in the Ba sample patterns correspond to the intense (203) reflections of the 6L-perovskite impurity phases described in the main text.  }
\end{figure}

Due to the lack of consistent diffraction data on Ba$_2$LuIrO$_6$, Sr$_2$LuIrO$_6$, Ba$_2$ScIrO$_6$, and Sr$_2$ScIrO$_6$, we decided to revisit their crystal structures with NPD as part of our systematic study. We also collected NPD data on a Ba$_2$YIrO$_6$ reference sample. Our main results are summarized in Fig.~\ref{Fig1}, where the NPD data with $T =$~4~K are shown as solid red squares for all five double perovskite iridates. The diffraction patterns for the three Ba samples look remarkably similar, while several additional weak peaks are present in the patterns for the two Sr samples. A small subset of these differences is highlighted in Fig.~\ref{Fig1}(f), which shows a zoomed-in version of the five diffraction patterns overplotted between 2$\theta =$~34 and 41$^\circ$. Below 2$\theta =$~39.5$^\circ$, only one prominent Bragg peak is observed for the Ba samples, while three peaks are observed for the two Sr samples. There are also some additional weak peaks in the Ba sample patterns, which can be indexed to a 6L-perovskite impurity phase with the same chemical composition that will be described in more detail below. The most intense impurity peak for each Ba sample pattern shown in Fig.~\ref{Fig1}(f) is labeled with a black arrow. 

\begin{table}[htb]
\begin{center}
\caption{Structural parameters for cubic Ba$_2B$IrO$_6$ ($B =$~Y, Lu, Sc) extracted from the refinements of the~1.54~\AA~neutron powder diffraction data. The lattice constants are in \AA. Note that the Ba, $B$, Ir, and O ions occupy the 8$c$, 4$b$, 4$a$, and 24$e$ Wyckoff sites respectively. } 

\begin{tabular}{l l l l l}
\hline 
\hline
Material & Ba$_2$YIrO$_6$ & Ba$_2$LuIrO$_6$ & Ba$_2$ScIrO$_6$ \\
\hline
$a$ & 8.3387(2) & 8.2843(2) & 8.1354(1) \\  
O $x$ & 0.2352(1) & 0.2376(2) & 0.2431(2) \\
R$_\mathrm{wp}$ & 5.44~\% & 5.65~\% & 4.34~\% \\ 
$\chi^2$ & 3.34 & 2.96 & 3.25 \\
\hline\hline
\end{tabular}
\end{center}
\end{table}

\begin{table}[htb]
\begin{center}
\caption{Structural parameters for monoclinic Sr$_2B$IrO$_6$ ($B =$~Lu, Sc) extracted from the refinements of the~1.54~\AA~neutron powder diffraction data. The lattice constants and bond distances are in \AA~and all angles are in degrees. Note that the Sr, $B$, Ir, and O ions occupy the 4$e$, 2$c$, 2$d$, and 4$e$ Wyckoff sites respectively.} 

\begin{tabular}{l l l l l}
\hline 
\hline
Material & Sr$_2$LuIrO$_6$ & Sr$_2$ScIrO$_6$ \\
\hline
$a$ & 5.7304(2) & 5.6463(2) \\  
$b$ & 5.7355(1) & 5.6227(1) \\
$c$ & 8.1063(2) & 7.9567(2) \\
$\beta$ & 90.090(5) & 90.080(5) \\
Sr $x$ & 0.4949(6) & 0.4937(8) \\ 
Sr $y$ & 0.5282(3) & 0.5153(4) \\ 
Sr $z$ & 0.2542(9) & 0.251(2) \\
O$_1$ $x$ & 0.2920(9) & 0.278(2) \\
O$_1$ $y$ & 0.272(1) & 0.269(2) \\
O$_1$ $z$ & 0.0338(7) & 0.024(1) \\
O$_2$ $x$ & 0.2279(8) & 0.237(1) \\
O$_2$ $y$ & 0.7958(9) & 0.775(2) \\
O$_2$ $z$ & 0.0337(8) & 0.027(1) \\
O$_3$ $x$ & 0.5661(8) & 0.5521(7) \\
O$_3$ $y$ & -0.0135(5) & -0.0097(7) \\
O$_3$ $z$ & 0.2355(6) & 0.248(2) \\
Site mixing & 3.4~\% & 4.4~\% \\
R$_\mathrm{wp}$ & 4.02~\% & 4.76~\% \\ 
$\chi^2$ & 2.22 & 3.30 \\
Ir-O$_1$ & 1.982(6) & 1.972(9) \\
Ir-O$_2$ & 1.969(5) & 1.965(9) \\ 
Ir-O$_3$ & 1.947(5) & 2.00(1) \\ 
O$_1$-Ir-O$_2$ &  91.6(4)  &  90.2(7)   \\ 
O$_2$-Ir-O$_3$ &  90.4(4)  &  90.8(7)    \\
O$_1$-Ir-O$_3$ &  90.7(4)  &  91.0(7)    \\
\hline\hline
\end{tabular}
\end{center}
\end{table}

Rietveld refinement results performed using FullProf \cite{93_rodriguez} and assuming stoichiometric chemical formulas are superimposed on the data shown in Fig.~\ref{Fig1} as black solid curves. Although Ir is volatile, the lack of significant impurity peaks associated with unreacted starting materials or intermediate phases shows that this was properly accounted for by adding extra Ir to the starting mixture used for the solid state synthesis. Tables~I and II show key refinement parameters for the Ba and Sr systems respectively. We find that the data for the three Ba systems refine best in the {\it Fm$\bar{3}$m} space group corresponding to the ideal FCC structure. In sharp contrast, the diffraction data for the two Sr materials are better described by the monoclinic {\it P$2_1$/n} space group. We can immediately rule out the non-cubic crystal field mechanism for generating magnetic order in four of these materials, as the IrO$_6$ octahedra are ideal in the three Ba systems and elongated due to an apical oxygen distortion in Sr$_2$ScIrO$_6$, which should generate a non-magnetic singlet state. The situation is different for Sr$_2$LuIrO$_6$, as the short iridium - apical oxygen distance suggests that the non-cubic crystal field mechanism may be viable. Regarding the excitonic condensation mechanism for generating magnetic order, it is notable that the lattice constants of Ba$_2$LuIrO$_6$ and Ba$_2$ScIrO$_6$ (Sr$_2$LuIrO$_6$ and Sr$_2$ScIrO$_6$) are significantly reduced compared to Ba$_2$YIrO$_6$ (Sr$_2$YIrO$_6$). The smaller unit cells will have shorter exchange pathways and therefore they should push these materials closer to the regime where excitonic magnetism is realized. 

Our analysis of the diffraction data also identified two potential sources of paramagnetic moments in these materials. While no $B/B'$ site mixing was found in the Ba samples, values of 3.4~\% and 4.4~\% were found for Sr$_2$LuIrO$_6$ and Sr$_2$ScIrO$_6$ respectively. This antisite disorder can create paramagnetic moments via local exciton condensation resulting from enhanced exchange interactions at neighboring $B$ and $B'$ sites \cite{17_chen}. We also found impurity peaks in our diffraction patterns for the Ba samples corresponding to the 6L-polytype Ba$_3 B (B_{0.25}$Ir$_{0.75})_2$O$_9$ crystallizing in the space group {\it P$\bar{3}$m1}. The 6L-perovskite impurity weight fractions were 7.3\%, 14.8\%, and 16.7\% for the Ba$_2$YIrO$_6$, Ba$_2$LuIrO$_6$, and Ba$_2$ScIrO$_6$ samples respectively. Notably, this impurity phase was identified previously in Ba$_2$ScIrO$_6$ \cite{17_kayser}, but not in Ba$_2$YIrO$_6$ or Ba$_2$LuIrO$_6$. The 6L-perovskite crystal structure consists of dimers arising from the mixed site with an extremely short intradimer distance \cite{17_ziat, 21_nguyen}, so metal-metal bonding is likely active for the dimers that consist of two Ir$^{5+}$ ions. This situation leads to quasi-molecular orbital formation with the 8 electrons per dimer site giving rise to a magnetic degree of freedom. Indeed, paramagnetic moments that arise from metal-metal bonding in the Ir dimers have been identified in the 6H-and 6L-perovskites Ba$_3 B$Ir$_2$O$_9$ ($B =$~Zn, Mg, Sr, Ca) \cite{16_nag, 18_nag_2}, with ordered ground states predicted by DFT for the Ca and Sr analogs \cite{18_nag_2}.

\section{IV. Single ion properties}

\begin{figure}
\centering
\scalebox{1}{\includegraphics{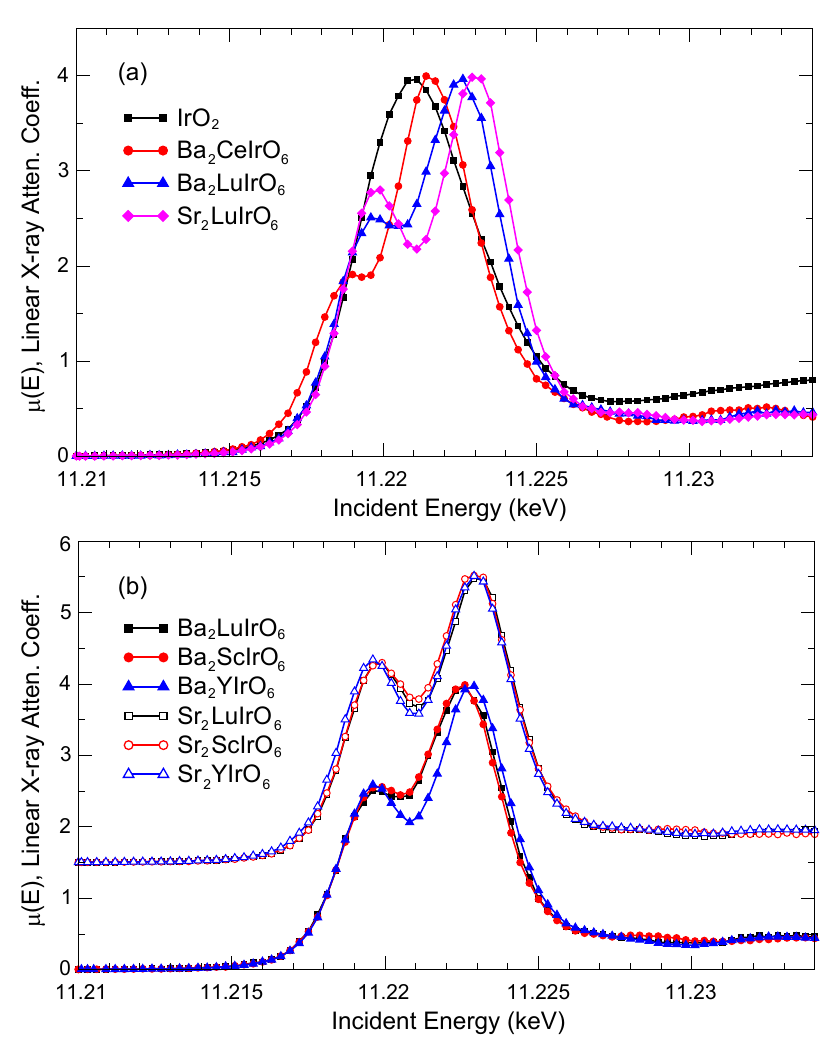}}
\caption{\label{Fig2} (color online) HERFD-XAS spectra collected at the Ir $L_3$ edge ($E_i$ = 11.215 keV) for selected 5$d^4$ double perovskite iridates and reference samples. (a) HERFD-XAS spectra from Ba$_2$LuIrO$_6$ and Sr$_2$LuIrO$_6$ compared to Ir$^{4+}$ reference samples IrO$_2$ and Ba$_2$CeIrO$_6$. The $\sim$1.2 eV chemical shift is consistent with a valence of Ir$^{5+}$. (b) HERFD-XAS spectra from Ba$_2$LuIrO$_6$, Sr$_2$LuIrO$_6$, Ba$_2$ScIrO$_6$, Sr$_2$ScIrO$_6$, Ba$_2$YIrO$_6$, and Sr$_2$YIrO$_6$. The spectra of the Sr-based compounds have been vertically offset for clarity.}
\end{figure}

High energy resolution fluorescence detected x-ray absorption spectroscopy (HERFD-XAS) can be used to extract important information regarding the valence of these compounds. Representative HERFD-XAS spectra for Ba$_2$LuIrO$_6$, Sr$_2$LuIrO$_6$, Ba$_2$ScIrO$_6$, and Sr$_2$ScIrO$_6$ are provided in Fig.~\ref{Fig2}, along with the spectra for a series of Ir$^{4+}$ (5$d^5$) and Ir$^{5+}$ (5$d^4$) reference samples. We note that all double perovskite iridates measured in this study display a characteristic two-peak absorption spectrum at the Ir $L_3$ edge. This is a sharp contrast to the one-peak Ir $L_3$ spectra observed for IrO$_2$ and other common Ir$^{3+}$ and Ir$^{4+}$ iridates reported elsewhere \cite{12_clancy}. This two-peak structure can be attributed to excitations from the 2$p_{3/2}$ core states to the unoccupied 5$d$ $t_{2g}$ states (lower energy peak) and $e_{g}$ states (higher energy peak). Although the cubic crystal field splitting (10$Dq$) between $t_{2g}$ and $e_g$ states in the Ir double perovskites is of similar magnitude to other iridates, the bandwidth of these states appears to be significantly narrower due to the greater separation between neighboring Ir ions. The HERFD-XAS spectra for all six 5$d^4$ double perovskite iridates shown in Fig.~\ref{Fig2}(b) are also clearly shifted to higher energies with respect to the Ir$^{4+}$ reference samples IrO$_2$ and Ba$_2$CeIrO$_6$. This chemical shift indicates that the Ir ions in these compounds are adopting the Ir$^{5+}$ (5$d^4$) oxidation state. We note that the presence of the 6L-perovskite impurity phases in the Ba samples does not have a significant effect on this data because their chemical compositions, and hence their expected Ir valence states, are the same as the double perovskite compounds of interest. 

\begin{table}[htb]
\begin{center}
\caption{HERFD-XAS fitting results for selected 5$d^4$ double perovskite iridates and 5$d^5$ reference samples (IrO$_2$ and Ba$_2$CeIrO$_6$). Parameters include the position of the white-line features associated with transitions into the Ir 5$d$ $t_{2g}$ ($E_{t2g}$) and 5$d$ $e_{g}$ ($E_{eg}$) states, the energy difference between these features (10$Dq$), the linewidth (FWHM), and the intensity ratio of the $e_{g}$ and $t_{2g}$ features. The $E_{t2g}$ and $E_{eg}$ parameters are listed in keV and the 10$Dq$ and FWHM parameters are given in eV.}
\begin{tabular}{l l l l l l}
\hline 
\hline
Material & $E_{t2g}$ & $E_{eg}$ & {10$Dq$} & FWHM & Int. Ratio \\
\hline
IrO$_2$ & --- & 11.2212(1) & --- & 3.8(4) & --- \\ 
Ba$_2$CeIrO$_6$ & 11.2183(1) & 11.2216(1) & 3.3(2) & 2.8(3) & 5.8(1.0) \\  
Ba$_2$LuIrO$_6$ & 11.2195(1) & 11.2226(1) & 3.1(2) & 2.6(3) & 2.0(3) \\ 
Ba$_2$ScIrO$_6$ & 11.2195(1) & 11.2225(1) & 3.0(2) & 2.6(3) & 1.9(3) \\
Ba$_2$YIrO$_6$ & 11.2195(1) & 11.2228(1) & 3.4(2) & 2.7(3) & 1.9(3) \\
Sr$_2$LuIrO$_6$ & 11.2197(1) & 11.2230(1) & 3.3(2) & 2.5(3) & 1.6(3) \\ 
Sr$_2$ScIrO$_6$ & 11.2197(1) & 11.2229(1) & 3.3(2) & 2.6(3) & 1.7(3) \\
Sr$_2$YIrO$_6$ & 11.2195(1) & 11.2230(1) & 3.5(2) & 2.7(3) & 1.7(3) \\
\hline\hline
\end{tabular}
\end{center}
\end{table}

\begin{figure*}
\centering
\scalebox{0.35}{\includegraphics{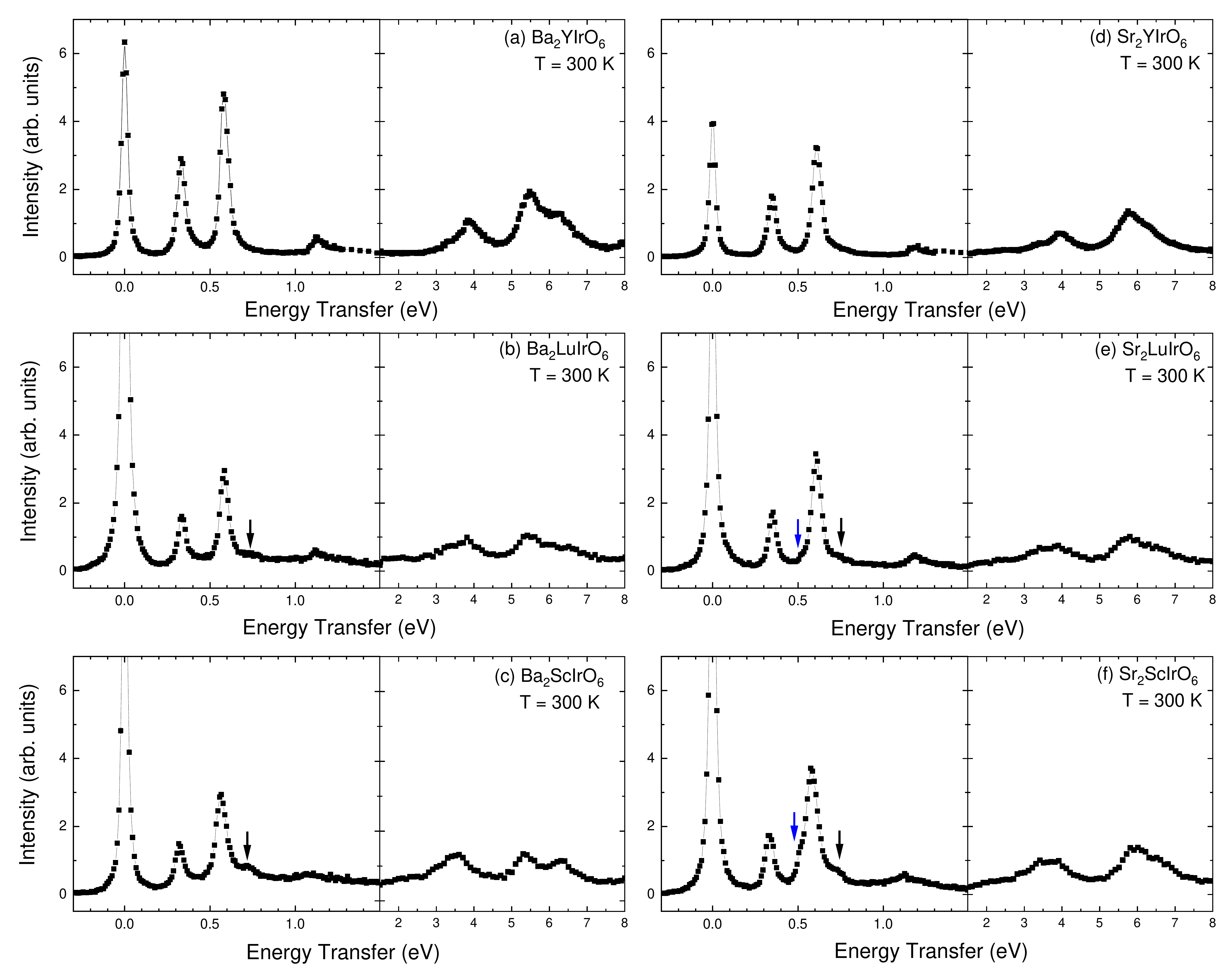}}
\caption{\label{Fig3} (color online) Resonant inelastic x-ray scattering spectra collected at the Ir $L_3$ edge ($E_i$ = 11.215 keV) for (a) Ba$_2$YIrO$_6$, (b) Ba$_2$LuIrO$_6$, (c) Ba$_2$ScIrO$_6$, (d) Sr$_2$YIrO$_6$, (e) Sr$_2$LuIrO$_6$, and (f) Sr$_2$ScIrO$_6$. Note the presence of a strong elastic line at an energy transfer $\hbar \omega$~$=$~0 and two prominent peaks between energy transfers of $\hbar \omega $~$\sim$ 0.2 and 0.8 eV corresponding to intra-$t_{2g}$ transitions associated with Ir$^{5+}$. There are also higher energy excitations between energy transfers of $\hbar \omega $~$\sim$2.5 and 5 eV corresponding to Ir$^{5+}$ interband $t_{2g}$ to $e_{g}$ transitions, and between $\sim$5 and 7.5 eV corresponding to charge-transfer excitations from the O 2$p$ band to the Ir 5$d$ bands. We attribute the features marked with the blue and black arrows to d-d excitations associated with Ir$^{6+}$ and Ir$^{4+}$ defects respectively. }
\end{figure*}

The strong white-line features at the Ir $L_3$ edge can be modeled using a simple phenomenological fit function which includes an arctangent function (to describe the step-like 2$p$ $\rightarrow$ continuum excitation process) and two Gaussian functions (to describe the sharp atomic-like 2$p$ $\rightarrow$ 5$d$ $t_{2g}$ and 2$p$ $\rightarrow$ 5$d$ $e_g$ excitation processes).  The key fit parameters extracted from this model are described in Table III. Although the HERFD-XAS spectra for the 5$d^4$ double perovskite iridates are extremely similar (Fig.~\ref{Fig2}b), the fitting analysis reveals subtle differences. As noted above, all six of the 5$d^4$ double perovskite iridates measured in this study display a $\sim$1.2 eV chemical shift compared to the Ir 5$d^5$ double perovskite reference sample. There is also no significant increase in peak width or peak splitting (10$Dq$) that might indicate mixed valence or the presence of substantial Ir$^{4+}$ or Ir$^{6+}$ defects. However, we do note a slight shift towards higher energies for Ba$_2$YIrO$_6$ and the three Sr-based compounds. Similarly, the intensity ratio of the two white-line features, which reflects the density of unoccupied 5d $t_{2g}$ and $e_g$ states, is slightly reduced in the Sr-based compounds. In an ideal 5$d^4$ system, one would expect 2 $t_{2g}$ holes and 4 $e_g$ holes, and a peak intensity ratio of 2. As a result, the slightly reduced intensity ratio is consistent with the presence of additional $t_{2g}$ holes, as would be produced by Ir$^{6+}$ (5$d^3$) defects. Within experimental uncertainty, all six of these compounds are consistent with a valence state of Ir$^{5+}$. The HERFD-XAS data suggests that any defect-induced effects are likely to be more pronounced in the Sr-based compounds, with a larger concentration of Ir$^{6+}$ defects (although some Ir$^{4+}$ defects may also be present).  

Resonant inelastic x-ray scattering (RIXS) provides a direct measurement of the spin-orbit coupling constant $\lambda_{\rm SOC}$ for 5$d^4$ iridates and therefore offers insight into the likelihood that they will exhibit magnetic order at low temperatures. Three or four intra-$t_{2g}$ excitations have been observed previously in the RIXS spectra of Ba$_2$YIrO$_6$ \cite{18_nag} and Sr$_2$YIrO$_6$ \cite{17_yuan}. The two lower energy peaks have a strong intensity because they arise from single-particle transitions across the spin-orbit gap 3$\lambda_{\rm SOC}$/2 that generate $J_{\rm eff} =$~1 and $J_{\rm eff} =$~2 states with an energy difference of Hund's coupling $J_{\rm H}$, while the higher-energy peaks correspond to two-particle excitations with weak intensities $\propto ~ (J_{\rm H}/\lambda_{\rm SOC})^2$ \cite{17_yuan}. Previous RIXS measurements on polycrystalline Ba$_2$YIrO$_6$ and Sr$_2$YIrO$_6$ have found singlet-triplet gap values of 350 meV \cite{18_nag} and 390 meV \cite{17_yuan} respectively. In both cases, the authors also diagonalized a local Hamiltonian with spin-orbit coupling and electron-electron interaction terms, which enabled a determination of the $\lambda_{\rm SOC}$ and $J_{\rm H}/\lambda_{\rm SOC}$ values required to simultaneously account for the three lowest energy excitations. Values of $\lambda_{\rm SOC} =$~390~meV and 420~meV were found for Ba$_2$YIrO$_6$ and Sr$_2$YIrO$_6$ respectively, with comparable $J_{\rm H}$ values of 250~meV for both systems. Theoretical calculations \cite{17_chen} and follow-up single crystal RIXS measurements \cite{18_kusch} also placed an upper bound of $\sim$~50~meV on the dispersion of the $J_{\rm eff} =$~1 excitations, which effectively rules out the excitonic condensation mechanism for generating magnetic order in these materials. Although RIXS cannot probe the strength of the non-cubic crystal field $\Delta$ at the Ir$^{5+}$ sites in 5$d^4$ systems, the non-cubic crystal field mechanism for generating magnetism is likely not operable in Sr$_2$YIrO$_6$ either, as RIXS measurements on 5$d^5$ double perovskite iridates with comparable IrO$_6$ octahedral distortions typically find similar $\lambda_{\rm SOC}$ values that are significantly larger than $\Delta$ \cite{19_aczel}. 

The RIXS spectra at the Ir L$_3$ edge for Ba$_2$LuIrO$_6$, Ba$_2$ScIrO$_6$, Sr$_2$LuIrO$_6$, and Sr$_2$ScIrO$_6$, as well as for reference samples of Ba$_2$YIrO$_6$ and Sr$_2$YIrO$_6$, are shown in Fig.~\ref{Fig3}. There are three inelastic modes observed below 1.5~eV for all samples with energy scales that are consistent with expectations for intra-$t_{2g}$ excitations associated with Ir$^{5+}$. There is also a series of modes between 2 and 8~eV with energy scales characteristic of interband $t_{2g}$ to $e_g$ transitions ($\sim$2.5 to 5 eV) and charge-transfer excitations from the O 2$p$ band to the Ir 5$d$ bands ($\sim$5 to 7.5 eV). For the remainder of this work, we will concentrate on the intra-$t_{2g}$ excitations. 

The two lowest-energy intra-$t_{2g}$ modes are much more intense and therefore likely arise from the single particle origin described above. We fit each spectrum below 1 eV to the sum of three pseudo-Voigt functions representing the elastic line and the two intra-$t_{2g}$ transitions found in this energy range. These results were used to check the accuracy of the elastic line position, establish an instrument energy resolution of 40~meV (full-width half-maximum), and determine precise inelastic peak positions ($\hbar \omega_1$ and $\hbar \omega_2$) to facilitate a meaningful quantitative comparison between all the samples. The fitting results are summarized in Table~IV. The procedure to estimate $\lambda_{\rm SOC}$ and $J_{\rm H}$ can be simplified through the realization that the peak positions for the two lower energy excitations are found at approximate energies of $3\lambda_{\rm SOC}/2 - J_{\rm H}$ and $3\lambda_{\rm SOC}/2$ \cite{17_yuan, 18_paramekanti, 18_nag}. The modified values of $\lambda_{\rm SOC}$ and $J_{\rm H}$ obtained for Ba$_2$YIrO$_6$ and Sr$_2$YIrO$_6$, based on the previously measured peak positions for the two lowest energy modes, are shown in Table~IV. Notably, the values determined by the two different procedures agree within the energy resolution of the MERIX spectrometer. 

\begin{table}[htb]
\begin{center}
\caption{RIXS fitting results of the intra-$t_{2g}$ excitations, spin-orbit coupling constants ($\lambda_{\rm SOC}$), and Hund's coupling $J_{\rm H}$ for selected 5$d^4$ double perovskite iridates. All parameters are in meV.} 
\begin{tabular}{l l l l l l}
\hline 
\hline
Material & $\hbar \omega_1$ & $\hbar \omega_2$ & $\lambda$ & $J_H$ & Ref. \\
\hline 
Ba$_2$YIrO$_6$ & 350 & 600 & 400 & 250 & \cite{18_nag} \\  
Ba$_2$YIrO$_6$ & 334(1) & 585(1) & 390(1) & 251(1) & this work \\
Sr$_2$YIrO$_6$ & 390(20) & 660(20) & 440(10) & 270(30) & \cite{17_yuan} \\
Sr$_2$YIrO$_6$ & 349(1) & 612(1) & 408(1) & 263(2) & this work \\
Ba$_2$LuIrO$_6$ & 337(2) & 585(1) & 390(1) & 248(2) & this work \\ 
Ba$_2$ScIrO$_6$ & 326(2) & 569(2) & 379(1) & 243(2) & this work   \\
Sr$_2$LuIrO$_6$ & 353(1) & 609(1) & 406(1) & 256(1) & this work \\ 
Sr$_2$ScIrO$_6$ & 335(1) & 582(1) & 388(1) & 247(2) & this work \\
\hline\hline
\end{tabular}
\end{center}
\end{table}

Our estimates of $\lambda_{\rm SOC}$ and $J_{\rm H}$ for all six 5$d^4$ double perovskite iridates, obtained using the simplified procedure described above, are also summarized in Table~IV. Ba$_2$ScIrO$_6$ has the lowest values for both $\lambda_{\rm SOC}$ and $J_{\rm H}$, with slightly larger values achieved by increasing the size of the $B$-site ionic radius and/or by inducing a structural distortion by substituting a smaller cation such as Sr on the $A$-site. Despite the small variation in $\lambda_{\rm SOC}$ and $J_{\rm H}$, the former parameter is large in all cases and therefore $J =$~0 electronic ground states are anticipated for this family of materials. 

We also find two small, additional inelastic features in the RIXS spectra at $\hbar \omega $~$\sim$~0.5 - 0.55 eV and 0.7 - 0.75 eV, as indicated by the blue and black arrows in Fig.~\ref{Fig3}. The latter feature is present in most of the spectra (possibly with the exception of the Ba$_2$YIrO$_6$ and Sr$_2$YIrO$_6$ data) and its energy scale is characteristic of an intra-$t_{2g}$ transition in 5$d^5$ double perovskite iridates \cite{19_aczel}, therefore we attribute it to small amounts of Ir$^{4+}$ defects in our samples. While the presence of Ir$^{4+}$ should also generate an excitation in the RIXS spectrum between 0.6 and 0.65~eV, this feature would exhibit strong overlap with one of the Ir$^{5+}$ modes. While the 0.5 - 0.55 eV peak is particularly prominent in the Sr$_2$ScIrO$_6$ spectrum and it may also be present in the Sr$_2$LuIrO$_6$ spectrum, it is noticeably absent in the Ba sample data. Although an unambiguous interpretation of this feature is hindered by the absence of a suitable Ir$^{6+}$ double perovskite reference sample, the HERFD-XAS data analysis presented in Table III suggests that this peak may be associated with a small Ir$^{6+}$ defect concentration.  Therefore, the combined HERFD-XAS and RIXS data can be explained by the presence of both Ir$^{4+}$ and Ir$^{6+}$ defects in these double perovskites, with more total defects in the Ba (or Sr) samples with higher final sintering temperatures and enhanced Ir$^{6+}$ defect concentrations in the Sr samples as compared to their analogous Ba sample counterparts. The simultaneous presence of Ir$^{4+}$ and Ir$^{6+}$ defects has been identified previously in Ba$_2$YIrO$_6$ via electron spin resonance measurements \cite{18_fuchs}. Finally, the 6L-perovskite impurities identified in the Ba samples should yield several broad peaks in the RIXS spectra between 0.2 and 0.8~meV \cite{19_nag}, but these features are likely too weak to be observed in the current data. We also note that the extra inelastic peaks that are discussed just above do not have the appropriate sample dependence to attribute to a 6L-perovskite impurity origin. 

\section{V. Search for magnetic order}

Ordered ground states were proposed previously for both Ba$_2$YIrO$_6$ \cite{14_cao} and Sr$_2$YIrO$_6$ \cite{17_terzic}, with common observations including sizable effective moments $\mu_{\rm eff} >$~0.6~$\mu_{\rm B}$ and large antiferromagnetic Curie-Weiss temperatures with $|\theta_{\rm CW}| >$~90~K. However, these results were not reproducible with other work typically finding $|\theta_{\rm CW}| <$~10~K and $\mu_{\rm eff} <$~0.6~$\mu_{\rm B}$ for these two materials and their solid solutions \cite{16_dey, 16_phelan, 17_chen, 17_corredor, 17_hammerath}. The wide range of values reported for the Curie-Weiss parameters of Ba$_2$YIrO$_6$ and Sr$_2$YIrO$_6$ suggest that they are determined primarily by extrinsic sources. Magnetic susceptibility studies of other 5$d^4$ double perovskite family members are less common, with a single report on Ba$_2$ScIrO$_6$ and Sr$_2$ScIrO$_6$ finding $\mu_{\rm eff}$ values of 0.39~$\mu_{\rm B}$ and 0.16~$\mu_{\rm B}$ respectively, $\theta_{\rm CW}$ values of 0~K, and no evidence of magnetic order in either compound \cite{17_kayser}. 

\begin{figure}
\centering
\scalebox{0.145}{\includegraphics{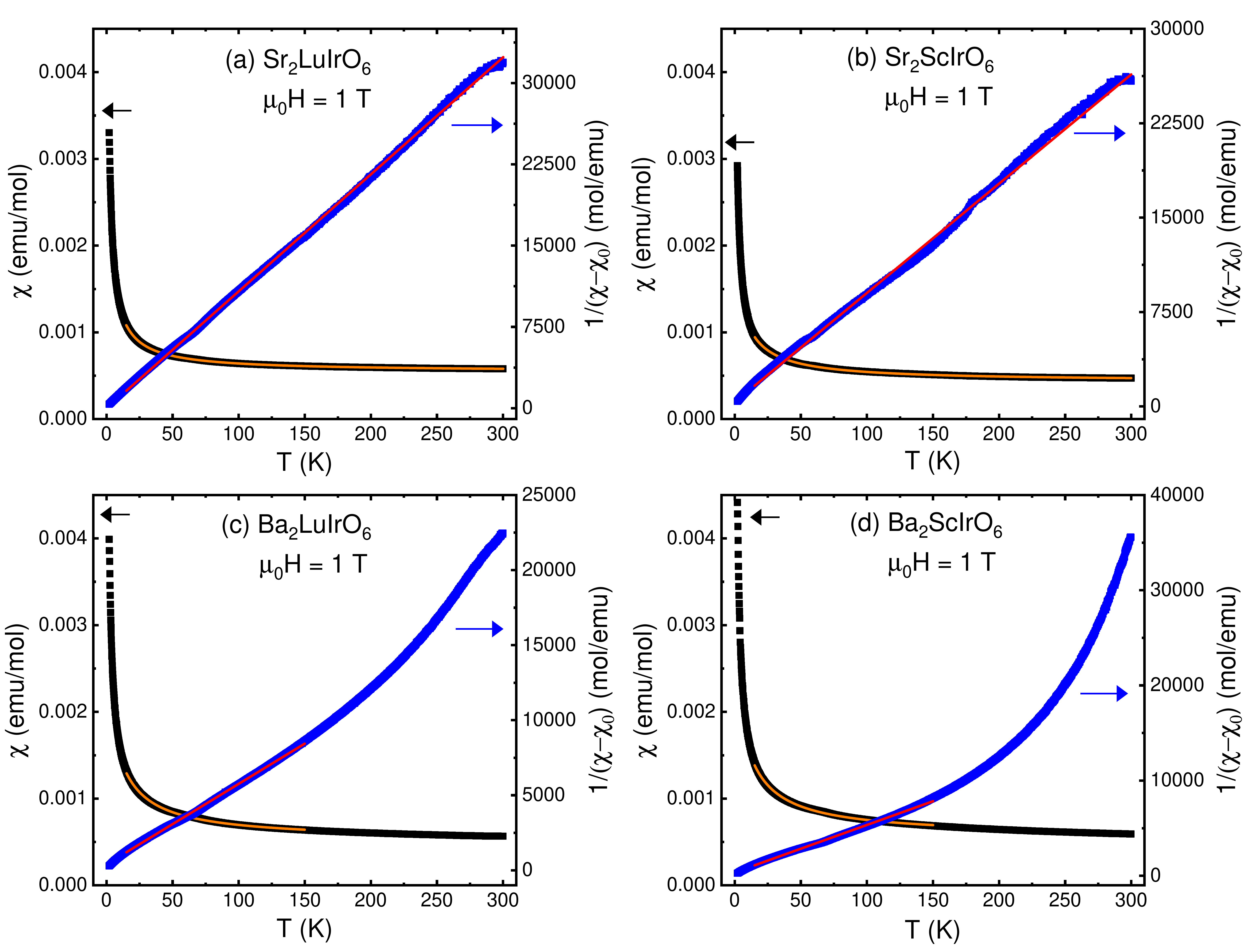}}
\caption{\label{Fig4} (color online) Temperature-dependence of the magnetic susceptibility (left axis) and inverse magnetic susceptibility (right axis) plotted as 1/($\chi - \chi_0$) in an applied magnetic field of 1~T for (a) Sr$_2$LuIrO$_6$, (b) Sr$_2$ScIrO$_6$, (c) Ba$_2$LuIrO$_6$, and (d) Ba$_2$ScIrO$_6$. The solid orange curves superimposed on the susceptibility data represent fits to a Curie-Weiss law plus a constant, while the solid red curves superimposed on the inverse susceptibility data show the linear fitting results.}
\end{figure}

Our magnetic susceptibility data for Sr$_2$LuIrO$_6$, Sr$_2$ScIrO$_6$, Ba$_2$LuIrO$_6$, and Ba$_2$ScIrO$_6$, plotted as $\chi = M/H$ in an applied magnetic field of 1 T, are presented in Fig.~\ref{Fig4}. There are no obvious signatures of magnetic ordering in this data. The magnetic signal from the Sr samples is well-explained by a Curie-Weiss law plus a constant $\chi_0$ over a wide temperature range (15 - 300 K), which is corroborated by the nearly-linear behavior of 1/($\chi - \chi_0$) shown in Fig.~\ref{Fig4}(a) and (b). These results agree well with several other 5d$^4$ double perovskite iridate studies \cite{16_dey, 17_chen, 17_corredor, 17_hammerath}. On the other hand, the parameters extracted from the Curie-Weiss law plus constant fits for the Ba samples are quite dependent on the choice of fitting range, which can be understood by the non-linear temperature-dependence of 1/($\chi - \chi_0$) depicted in Fig.~\ref{Fig4}(c) and (d). To facilitate a quantitative comparison with the Sr sample data, we employed an iterative three-step procedure for fitting the Ba sample data. We first fit the susceptibility data to a Curie-Weiss law plus a constant over a particular temperature range 15 K - $T_{\rm max}$, then plotted 1/($\chi - \chi_0$), and finally fit the 1/($\chi - \chi_0$) to a simple linear function to check if this $T$-dependence explained the data well over the entire fitting range. If non-linear behavior was observed in the higher-temperature regime, we repeated the procedure with a lower value for $T_{\rm max}$. We ultimately settled on the fitting window of 15 - 150 K for the Ba samples where 1/($\chi - \chi_0$) exhibits linear behavior. The fitting parameters for all four systems are summarized in Table~V. We find large, positive $\chi_0$ values on the order of 10$^{-4}$ emu/mol indicative of Van Vleck paramagnetism and effective moments $\mu_{\rm eff} <$~0.6~$\mu_{\rm B}$, which are in broad agreement with most reported results for the 5$d^4$ double perovskite iridates. We have also included upper bound estimates for the Ir$^{4+}$ and Ir$^{6+}$ defect concentrations in Table~V by assuming that the Curie-Weiss behavior was entirely associated with each defect type in turn. These estimates also assume that the Ir$^{4+}$ and Ir$^{6+}$ defects generate spin-1/2 and spin-3/2 moments respectively. 

\begin{table}[htb]
\begin{center}
\caption{Magnetic susceptibility fitting results to a Curie-Weiss law plus a constant for various 5$d^4$ double perovskite iridates.} 
\begin{tabular}{l l l l l l}
\hline 
\hline
Material & Ba$_2$LuIrO$_6$ & Ba$_2$ScIrO$_6$ & Sr$_2$LuIrO$_6$ & Sr$_2$ScIrO$_6$ \\
\hline
Fitting range (K) & 15 - 150 & 15 - 150 & 15 - 300 & 15 - 300 \\
$\mu_{\rm eff}$ ($\mu_{\rm B}$) & 0.39(1) & 0.41(1) & 0.28(1) & 0.31(1) \\
$\theta_{\rm CW}$ (K) & -9.9(1) & -10.5(2) & -2.89(2) & -8.1(1) \\
$\chi_0$ (10$^{-4}$ emu/mol) & 5.20(1) & 5.64(1) & 5.48(1) & 4.33(1) \\
Ir$^{4+}$ defects & $<$ 5.1\% & $<$ 5.6\% & $<$ 2.6\% & $<$ 3.2\% \\
Ir$^{6+}$ defects & $<$ 1.0\% & $<$ 1.1\% & $<$ 0.5\% & $<$ 0.6\% \\
\hline\hline
\end{tabular}
\end{center}
\end{table}

\begin{figure*}
\centering
\scalebox{0.28}{\includegraphics{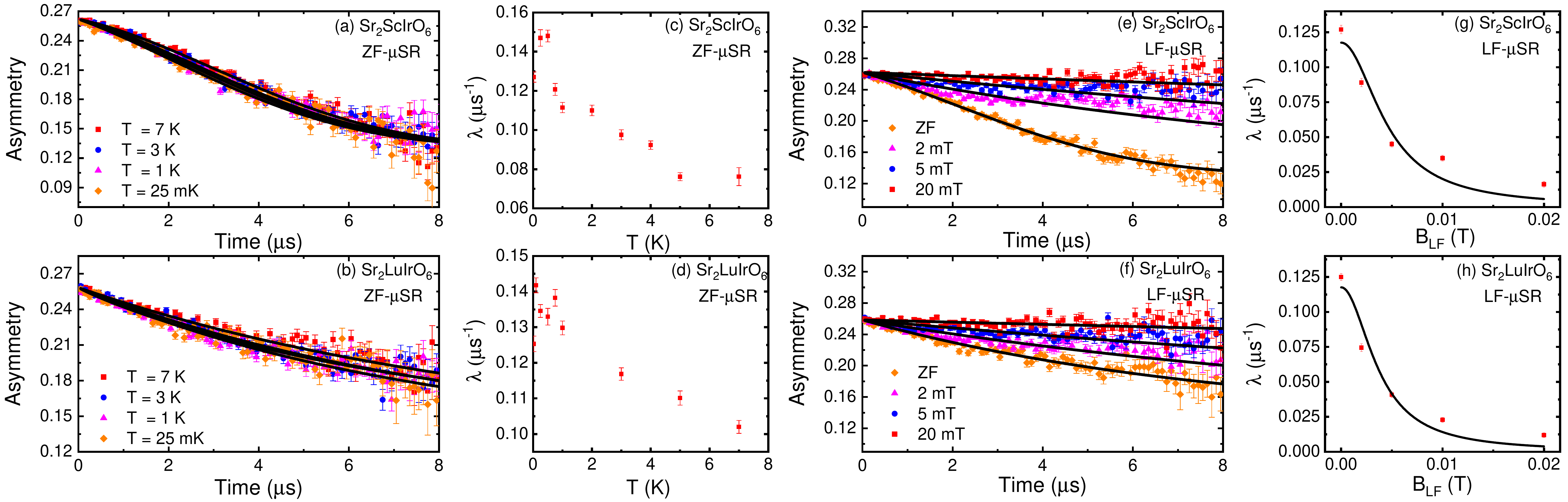}}
\caption{\label{Fig5} (color online) Zero-field muon asymmetry vs time at selected temperatures for (a) Sr$_2$ScIrO$_6$ and (b) Sr$_2$LuIrO$_6$. The black curves superimposed on the data represent fits to the functional form described in the main text. Key zero-field fitting parameters for (c) Sr$_2$ScIrO$_6$ and (d) Sr$_2$LuIrO$_6$. Longitudinal-field muon asymmetry vs time at selected temperatures for (e) Sr$_2$ScIrO$_6$ and (f) Sr$_2$LuIrO$_6$. The black curves superimposed on the data again represent fits to the functional form described in the main text. Longitudinal-field relaxation rates for (g) Sr$_2$ScIrO$_6$ and (h) Sr$_2$LuIrO$_6$, with Redfield theory fits superimposed on the data. }
\end{figure*}

Since our RIXS results identified large $\lambda_{SOC}$ values between 379 and 408~meV for the 5$d^4$ double perovskite iridates considered here, we expect the formation of a non-magnetic singlet ground state (i.e. a Van Vleck paramagnet) that would be characterized by a temperature-independent susceptibility over a wide temperature range. However, even the susceptibility data for the Sr samples shows a significant deviation from this behavior. The presence of local moments in Sr$_2$LuIrO$_6$ and Sr$_2$ScIrO$_6$ likely arises from the extrinsic sources identified above including the antisite disorder and the finite Ir$^{4+}$ and Ir$^{6+}$ defect concentrations, but intrinsic origins cannot be completely ruled out on the basis of these measurements alone. The susceptibility data for Ba$_2$LuIrO$_6$ and Ba$_2$ScIrO$_6$ are even more challenging to interpret, as the significant 6L-perovskite impurity phases identified in the neutron diffraction measurements described above generate a more complicated temperature-dependence. The deviations of the 1/($\chi - \chi_0$) data from simple linear behavior may arise from a subtle magnetic ordering signal from the 6L-perovskite impurity phases, but the precise ordering temperature is difficult to pinpoint. Another possibility is that this deviation arises from a magnetic susceptibility contribution associated with a molecular d-d excitation of the Ir dimers in the 6L-perovskite impurity phases, as identified previously in related 6H-perovskites based on heavy transition metal dimers \cite{17_ziat}. This latter explanation seems more likely as then the enhanced Curie contributions in the Ba samples could arise from effective moments associated with both Ir$^{4+}$/Ir$^{6+}$ defects and the paramagnetic 6L-perovskite impurity phases. We note that attributing the enhanced Curie constants in the Ba samples to Ir$^{4+}$/Ir$^{6+}$ defects only is not consistent with their smaller combined concentrations as inferred from our RIXS data. In any case, these discrepancies from the expectations for Van Vleck paramagnetism warrant further investigation. 

Muon spin relaxation is the ideal technique for establishing the magnetic ground states of the 5$d^4$ double perovskite iridates because it can detect spin freezing and ordered spin configurations with extremely small moments. Notably, the measured frequency in a zero-field muon spin relaxation (ZF-$\mu$SR) experiment is directly proportional to the ordered moment size in the material and depolarization of the asymmetry spectrum can arise from static mechanisms (i.e. local field inhomogenities) or dynamic mechanisms (i.e. spin fluctuations). $\mu$SR is also a real-space, local probe of magnetism that is sensitive to ordered volume fractions, so it can be used to identify magnetic impurity contributions that may be difficult to disentangle from intrinsic signals using magnetic susceptibility data only. The time-evolution of the muon spin polarization determined from a ZF-$\mu$SR measurement, plotted as the muon asymmetry \cite{97_dereotier, yaouanc_textbook}, is shown in Fig.~\ref{Fig5}(a) and (b) for Sr$_2$ScIrO$_6$ and Sr$_2$LuIrO$_6$ respectively. Quick inspection of the data reveals no evidence for the oscillations that represent one of the characteristic features of static magnetism in a polycrystalline sample. Instead, the weakly-relaxing spectra fit well to the following functional form: 
\begin{equation}
A(t) = A_s e^{-\lambda t} \times G_{KT}(t) + A_{\rm BKG},
\end{equation}
where $A_s$ is the initial asymmetry of the sample, $\lambda$ is the electronic relaxation rate, and $A_{\rm BKG}$ is the background asymmetry associated with the Ag sample holder and Ag powder mixed in with the samples. The fitting results are shown as solid curves superimposed on the data in Fig.~\ref{Fig5}(a) and (b). The Gaussian Kubo-Toyabe function of the form:
\begin{equation}
G_{KT}(t) = \frac{1}{3} + \frac{2}{3}(1-\sigma^2 t^2) e^\frac{-\sigma^2 t^2}{2} 
\end{equation}
was only required for Sr$_2$ScIrO$_6$ (i.e. it was set to 1 for Sr$_2$LuIrO$_6$) to account for the sizable quasi-static nuclear moments associated with $^{45}$Sc. In fact, this temperature-independent contribution dominates the Sr$_2$ScIrO$_6$ spectra. To estimate the volume fractions for the Sr$_2$ScIrO$_6$ sample and the Ag contribution, we first obtained the total asymmetry ($A_{\rm tot} = A_s + A_{\rm BKG}$) from a weak transverse field measurement at 7~K and then we fit the 7~K zero-field spectrum to Eq.~(2) with this constraint imposed. Our fit yields $\sigma =$~0.17~$\mu s^{-1}$ and $A_s = A_{\rm BKG}$. The former value is a typical relaxation rate for strong nuclear moments in $\mu$SR and a 50\% sample volume fraction is reasonable based on the relative sizes of the sample and the Ag sample holder and the fact that additional Ag powder was added to the sample pellet to improve thermal conductivity in the experiment. Since the size and Ag content of both Sr samples used in this experiment were nearly identical, we imposed the constraint $A_s = A_{\rm BKG}$ in the Sr$_2$LuIrO$_6$ data fitting.

\begin{figure*}
\centering
\scalebox{0.28}{\includegraphics{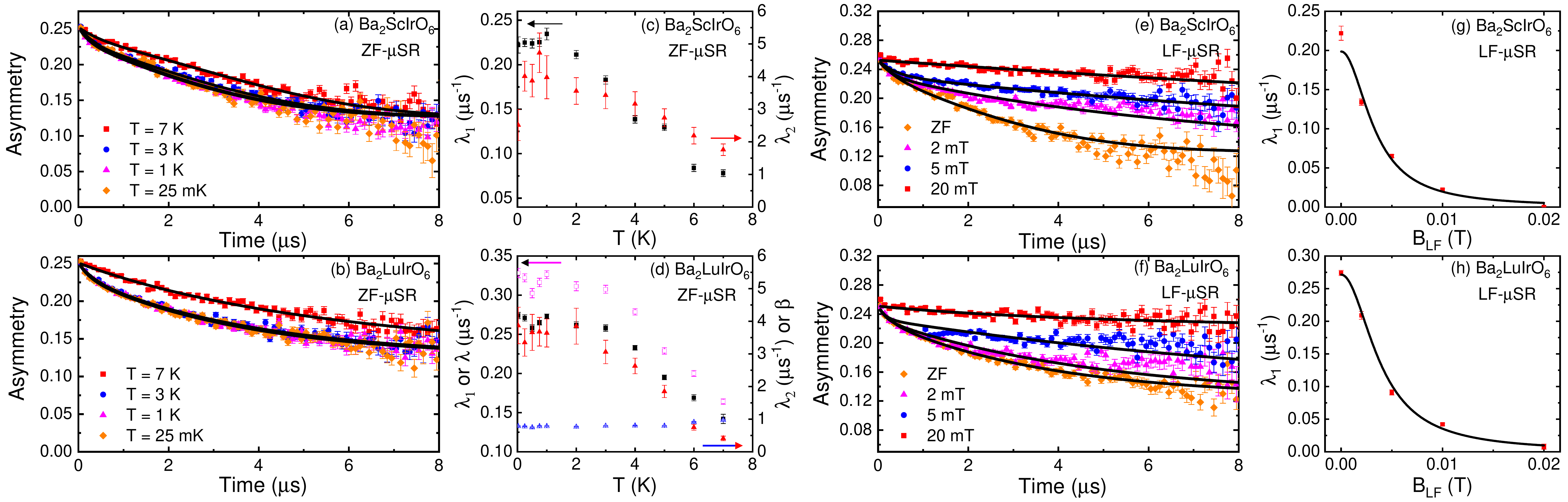}}
\caption{\label{Fig6} (color online)  Zero-field muon asymmetry vs time at selected temperatures for (a) Ba$_2$ScIrO$_6$ and (b) Ba$_2$LuIrO$_6$. The black curves superimposed on the data represent fits to the functional form with two sample components described in the main text. Key zero-field fitting parameters for (c) Ba$_2$ScIrO$_6$ and (d) Ba$_2$LuIrO$_6$. For the Lu sample, the fit parameters are shown for both the two sample component and power exponential functions described in the main text. Longitudinal-field muon asymmetry vs time at selected temperatures for (e) Ba$_2$ScIrO$_6$ and (f) Ba$_2$LuIrO$_6$. The black curves superimposed on the data again represent fits to the functional form with two sample components described in the main text. Longitudinal-field relaxation rates for (g) Ba$_2$ScIrO$_6$ and (h) Ba$_2$LuIrO$_6$, with Redfield theory fits superimposed on the data.  }
\end{figure*}

The temperature-dependence of the electronic moment relaxation rate $\lambda$ is shown in Fig.~\ref{Fig5}(c) and (d) for Sr$_2$ScIrO$_6$ and Sr$_2$LuIrO$_6$ respectively. While $\lambda$ increases weakly with decreasing temperature, there is no sign of the characteristic peak that would be expected for spin freezing. The single relaxation component also rules out magnetic order with a broadened local field distribution due to crystallographic defects or domain walls, as one expects a polycrystalline sample spectrum with a 2/3 fast-relaxing component and a 1/3 slowly-relaxing tail in this case \cite{17_ziat}. The exponential form of the relaxation for the electronic moments is consistent with two common scenarios: (1) a dense magnetic system with fast fluctuating moments, or (2) a dilute magnetic system with static or slow-to-intermediate fluctuating moments \cite{uemura_book}. To differentiate between these two possibilities, we performed longitudinal field muon spin relaxation (LF-$\mu$SR). Some selected asymmetry spectra for both Sr$_2$ScIrO$_6$ and Sr$_2$LuIrO$_6$ with $T =$~25~mK at various longitudinal fields $B_{LF}$ are shown in Fig.~\ref{Fig5}(e) and (f). In the static limit, the zero-field relaxation should be completely decoupled when $B_{LF} = $~10$B_s$ \cite{uemura_book}, where $B_s$ is the average static local field at the muon site. $B_s$ can be estimated by the expressions $B_s = \sigma/\gamma_\mu$ or $B_s = \lambda/\gamma_\mu$, where $\sigma$ is the nuclear moment relaxation rate, $\lambda$ is the electronic moment relaxation rate and $\gamma_\mu =$~2$\pi \times$135.5~MHz/T is the muon gyromagnetic ratio. The $B_s$ value for the nuclear moments of Sr$_2$ScIrO$_6$ is 0.2~mT and the asymmetry spectrum completely loses its Gaussian character by $B_{LF} =$~2~mT, which confirms that the nuclear moments are static on the $\mu$SR timescale. On the other hand, since the $B_s$ values for the electronic moments of Sr$_2$LuIrO$_6$ and Sr$_2$ScIrO$_6$ at 25~mK are 0.15~mT and complete decoupling of the spectra does not occur until $B_{LF} =$~20~mT, static magnetism associated with them can be ruled out. Differentiating between the two dynamic magnetism scenarios is more difficult. We attempt to do this by first fitting the LF-$\mu$SR spectra to a modified version of Eq.~(2) with $G_{KT}(t)$ set to 1 and then use Redfield theory \cite{Slichter} to model the field-dependence of the relaxation rate according to the following expression: 
\begin{equation}
\lambda = \frac{1}{T_1} = \frac{2 \gamma_\mu^2 B_D^2 \omega_c}{\gamma_\mu^2 B_{LF}^2 + \omega_c^2}, 
\label{Redfield}
\end{equation}
where 1/$T_1$ is the spin-lattice relaxation rate, $B_D$ is the magnitude of the fluctuating field, and $\omega_c$ is a single characteristic fluctuation rate. The LF-$\mu$SR fitting results are shown as solid curves superimposed on the data in Fig.~\ref{Fig5}(e) and (f), while the $B_{LF}$-dependence of the electronic relaxation rates and the Redfield theory fits are shown in Fig.~\ref{Fig5}(g) and (h). The $B_D$ and $\omega_c$ values for Sr$_2$LuIrO$_6$ are 3.1(6)~MHz and 0.50(4)~mT and for Sr$_2$ScIrO$_6$ are 3.9(9)~MHz and 0.56(5)~mT, which are comparable to the values of 1.2~MHz and 0.2~mT found previously for Ba$_2$YIrO$_6$ \cite{18_nag}. The $\omega_c/\gamma_\mu B_D$ ratios are 7 and 8 for the Lu and Sc systems respectively, which places them on the border between the intermediate and fast fluctuating ($\omega_c/\gamma_\mu B_D >>$ 1) regimes. However, a system with dense magnetic Ir$^{5+}$ moments seems quite unlikely due to the dynamic local field strengths that are comparable to the static local fields of 0.2~mT arising from the tiny, quasi-static $^{45}$Sc nuclear moments. Therefore, the $\mu$SR data for Sr$_2$LuIrO$_6$ and Sr$_2$ScIrO$_6$ are best explained by dilute electronic spins that slow down with decreasing temperature. These local moments are generated by a combination of the 3-5\% level of antisite disorder determined by our NPD refinements and the small concentrations of Ir$^{4+}$ and Ir$^{6+}$ defects identified in our RIXS and magnetic susceptibility results. The increase in relaxation rate with decreasing temperature may arise from correlated defect or antisite disorder clusters, as proposed in previous work \cite{17_corredor, 18_fuchs}, leading to a competition between cluster exchange and entropy to minimize the Helmholtz free energy. 

The ZF-$\mu$SR data for Ba$_2$ScIrO$_6$ and Ba$_2$LuIrO$_6$ is complicated by the presence of the 6L-perovskite impurity phase as identified by NPD. Some selected ZF-$\mu$SR spectra for these two samples is plotted in Fig.~\ref{Fig6}(a) and (b). The spectra are qualitatively different as compared to the Sr samples due to the appearance of a fast-relaxing component with a small amplitude, which is best observed for Ba$_2$ScIrO$_6$ - the material known to have the largest 6L-perovskite impurity. To account for this additional feature, the data were fit to the following functional form: 
\begin{equation}
A(t) = A_{s1} e^{-\lambda_1 t} \times G_{KT}(t) + A_{s2} e^{-\lambda_2 t} + A_{\rm BKG}
\end{equation}
Similar to the case with the Sr samples, careful inspection of the asymmetry data reveal that the Ag background comprises about half of the measured signal. In our fits, we therefore used a modified constraint $A_{s1} + A_{s2} = A_{\rm BKG}$, where $A_{s2}$ was set to the 6L-perovskite impurity phase fraction as determined by our NPD measurements. The Gaussian Kubo-Toyabe function was again needed for Ba$_2$ScIrO$_6$ to account for the sizable $^{45}$Sc quasi-static nuclear moments, with a slightly higher $\sigma$ value of 0.18~$\mu s^{-1}$ due to the small additional Ba nuclei contributions. The fitting results are presented in Fig.~\ref{Fig6}(a) and (b), where it can be seen that the two-component relaxation function describes the data quite well. 

The temperature-dependence of the two electronic relaxation rates $\lambda_1$ and $\lambda_2$ is illustrated in Fig.~\ref{Fig6}(c) for Ba$_2$ScIrO$_6$ and (d) for Ba$_2$LuIrO$_6$. The fast-relaxing front-end of the spectrum arises from the 6L-perovskite impurity phase, while the slow-relaxing component is intrinsic to the samples of interest. We present LF-$\mu$SR measurements in Fig.~\ref{Fig6}(e) and (f), which were used to help identify the mechanism for the different relaxation components. For Ba$_2$ScIrO$_6$, the asymmetry spectrum again completely loses its Gaussian character by $B_{LF} =$~2~mT ($\approx 10 B_s$), so the nuclear moments exhibit static magnetism. The fast exponential relaxation rates in the 25~mK zero-field spectra provide static local field estimates of $B_s =$~3~mT and 4.5~mT for the Sc and Lu systems respectively. Since the LF-$\mu$SR spectra of both samples show complete decoupling when $B_{LF} =$~20~mT ($< 10 B_s$), this suggests that the 6L-perovskite impurity phases host some form of static magnetism. On the other hand, the decoupling of the slow exponential relaxation rate in the LF-$\mu$SR measurements proceeds in a nearly identical manner to our findings for the Sr samples. To quantify this behavior, we fit the LF-$\mu$SR data to a modified version of Eq.~(5) with $G_{KT}(t)$ set to 1 and then used Redfield theory to model the field-dependence of the slow exponential relaxation rate. The fitting results are shown as solid curves superimposed on the data in Fig.~\ref{Fig6}(e) and (f), while the Redfield theory results are shown in Fig.~\ref{Fig6}(g) and (h). We find $\omega_c =$~2.8(4)~MHz and $B_D =$~0.62(3)~mT for Ba$_2$ScIrO$_6$ and similar values of 3.3(0.2)~MHz and 0.78(2)~mT for Ba$_2$LuIrO$_6$. The $\omega_c/\gamma_\mu B_D$ ratios of 5 for both samples are comparable to the values obtained for the Sr systems above. The similar decoupling behavior for the intrinsic relaxation component of all four samples points to a common dominant origin for this depolarization, which we attribute to a combination of dilute magnetic Ir$^{4+}$ and Ir$^{6+}$ defects since no antisite disorder was found in the Ba samples with neutron powder diffraction. 

\begin{figure}
\centering
\scalebox{0.29}{\includegraphics{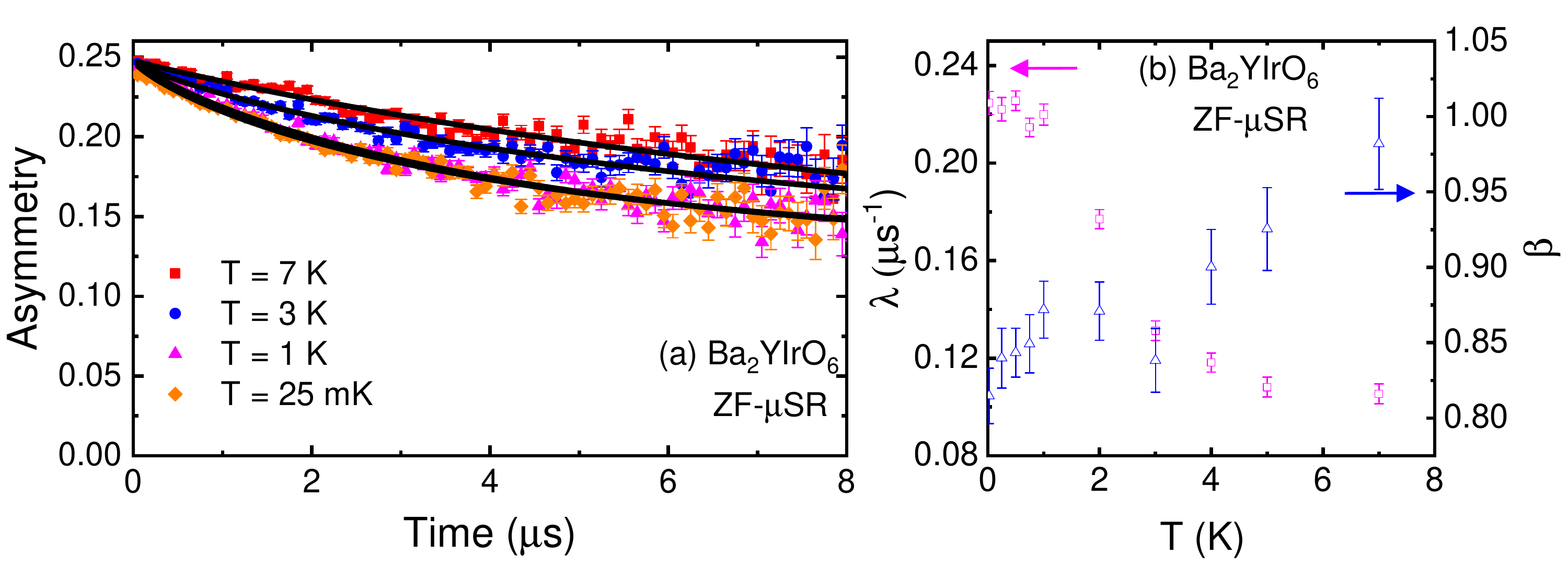}}
\caption{\label{Fig7} (color online) (a) Zero-field muon asymmetry vs time at selected temperatures for Ba$_2$YIrO$_6$. The black curves superimposed on the data represent fits to the functional form described in the main text. (b) Key zero-field fitting parameters for Ba$_2$YIrO$_6$. }
\end{figure}

Finally, we note that previous ZF-$\mu$SR measurements on the isostructural 5$d^4$ double perovskite Ba$_2$YIrO$_6$ \cite{17_hammerath, 18_nag} have been modelled with a single power exponential function rather than the two-component sample function we have employed here for our ZF-$\mu$SR data analysis of Ba$_2$ScIrO$_6$ and Ba$_2$LuIrO$_6$. Motivated by this difference, we also collected ZF-$\mu$SR data on our Ba$_2$YIrO$_6$ sample with the known 6L-perovskite impurity phase. The data at selected temperatures is shown in Fig.~\ref{Fig7}(a). While we first attempted to fit the data to the function with two sample components given by Eq.~(5), the fits did not converge for the lowest-temperature spectra. We attribute this issue to the significantly reduced volume fraction of the 6L-perovskite impurity phase in our Ba$_2$YIrO$_6$ sample (relative to our Ba$_2$LuIrO$_6$ and Ba$_2$ScIrO$_6$ samples), which leads to a fast-relaxing component with an extremely small amplitude. On the other hand, a power exponential function of the form:
\begin{equation}
A(t) = A_s e^{-(\lambda t)^\beta} + A_{\rm BKG},
\end{equation}
explains the ZF-$\mu$SR data quite well, as illustrated by the solid curves superimposed on the data in Fig.~\ref{Fig7}(a), which is in good agreement with previous work \cite{17_hammerath, 18_nag}. Fig.~\ref{Fig7}(b) presents the temperature-dependence of the fit parameters $\lambda$ and $\beta$. As a sanity check, we have also tried to fit our ZF spectra for Ba$_2$LuIrO$_6$ to the same power exponential function, as the presence of the fast relaxing component is not quite as obvious when compared to the Ba$_2$ScIrO$_6$ data. We find that the power exponential function fits also explain these data reasonably well, as the fast-relaxing front end is accounted for by a power $\beta <$~1. The fitting parameters, $\lambda$ and $\beta$, are shown as the magneta and blue open symbols in Fig.~\ref{Fig6}(d). The success of the power exponential function in describing our Ba$_2$YIrO$_6$ ZF-$\mu$SR data, along with its ability to adequately describe the Ba$_2$LuIrO$_6$ ZF-spectra, suggest that the non-exponential relaxation identified in the previous Ba$_2$YIrO$_6$  ZF-$\mu$SR data may arise from a small amount of the same 6L-perovskite impurity phase. 

\section{VI. Conclusions}
We performed a systematic study investigating the crystal structures, single ion properties, and magnetic ground states of Ba$_2$LuIrO$_6$, Ba$_2$ScIrO$_6$, Sr$_2$LuIrO$_6$, and Sr$_2$ScIrO$_6$. We found that the Ba systems crystallize in the cubic space group {\it Fm$\bar{3}$m}, while the Sr analogs form the {\it P$2_1$/n} structure. Despite the variation in the crystal structures, the Ir$^{5+}$ single ion properties are remarkably similar with comparable values for the spin-orbit coupling constant $\lambda_{\rm SOC}$ (379 - 406 meV) and Hund's coupling $J_{\rm H}$ (243 - 256 meV). Combined magnetic susceptibility and muon spin relaxation results reveal no evidence for magnetic order or spin freezing down to 25~mK in this family of materials. Our results are indicative of $J =$~0 electronic ground states in the 5$d^4$ double perovskite iridates due to large spin-orbit coupling, which renders the non-cubic crystal field and excitonic condensation mechanisms for generating magnetic order inactive. The local magnetic moments identified in these materials likely arise from extrinsic sources, and may be attributed to the 6L-perovskite impurity phases, the antisite disorder, and/or the Ir$^{4+}$ or Ir$^{6+}$ defects found in this work. We expect our findings to be broadly applicable to materials with octahedrally-coordinated transition metals in a 5$d^4$ electronic configuration.  

\section{acknowledgments}

A portion of this research used resources at the High Flux Isotope Reactor, which is a DOE Office of Science User Facility operated by Oak Ridge National Laboratory. This work is based upon research conducted at the Center for High Energy X-ray Sciences (CHEXS) which is supported by the National Science Foundation under award DMR-1829070. This research used resources of the Advanced Photon Source, a US Department of Energy (DOE) Office of Science User Facility operated for the DOE Office of Science by Argonne National Laboratory under contract No. DE-AC02-06CH11357.  Research at the University of Tennessee and the University of Illinois was supported by the National Science Foundation, Division of Materials Research under awards DMR-1455264 (G.J.M) and DMR-2003117 (H.D.Z). Work at McMaster University was supported by the Natural Sciences and Engineering Research Council of Canada. N.L. was supported by a Niedbala Family Fellowship at Villanova University.


\begin{thebibliography}{99}
\bibitem{14_witczak} W. Witczak-Krempa, G. Chen, Y.B. Kim, and L. Balents, Annu. Rev. Condens. Matter Phys. {\bf 5}, 57 (2014). 
\bibitem{16_du} Y. Du and X. Wan, Computational Materials Science {\bf 112}, 416 (2016). 
\bibitem{16_rau} J.G. Rau, E. Kin-Ho Lee, and H.-Y. Kee, Annu. Rev. Condens. Matter Phys. {\bf 7}, 195 (2016). 
\bibitem{16_schaffer} R. Schaffer, E. Kin-Ho Lee, B.-J. Yang, and Y.B. Kim, Rep. Prog. Phys. {\bf 79}, 094504 (2016). 
\bibitem{17_martins} C. Martins, M. Aichhorn, and S. Biermann, J. Phys.: Condens. Matter {\bf 29}, 263001 (2017). 
\bibitem{18_cao} G. Cao and P. Schlottmann, Rep. Prog. Phys. {\bf 81}, 042502 (2018). 
\bibitem{21_takayama} T. Takayama, J. Chaloupka, A. Smerald, G. Khaliullin, and H. Takagi, J. Phys. Soc. Jpn. {\bf 90}, 062001 (2021). 
\bibitem{08_kim} B.J. Kim, H. Jin, S.J. Moon, J.-Y. Kim, B.-G. Park, C.S. Leem, J. Yu, T.W. Noh, C. Kim, S.-J. Oh, J.-H. Park, V. Durairaj, G. Cao, and E. Rotenberg, Phys. Rev. Lett. {\bf 101}, 076402 (2008).
\bibitem{16_banerjee} A. Banerjee, C.A. Bridges, J.-Q. Yan, A.A. Aczel, L. Li, M.B. Stone, G.E. Granroth, M.D. Lumsden, Y. Yiu, J. Knolle, S. Bhattacharjee, D.L. Kovrizhin, R. Moessner, D.A. Tennant, D.G. Mandrus, and S.E. Nagler, Nature Materials {\bf 15}, 733 (2016).
\bibitem{17_banerjee} A. Banerjee, J.-Q. Yan, J. Knolle, C.A. Bridges, M.B. Stone, M.D. Lumsden, D.G. Mandrus, D.A. Tennant, R. Moessner, and S.E. Nagler, Science {\bf 356}, 1055 (2017).
\bibitem{17_winter} S.M. Winter, A.A. Tsirlin, M. Daghofer, J. van den Brink, Y. Singh, P. Gegenwart, and R. Valenti, J. Phys.: Condens. Matter {\bf 29}, 493002 (2017).
\bibitem{18_hermanns} M. Hermanns, I. Kimchi, and J. Knolle, Annu. Rev. Condens. Matter Phys. {\bf 9}, 17 (2018). 
\bibitem{19_takagi} H. Takagi, T. Takayama, G. Jackeli, G. Khaliullin, and S.E. Nagler, Nature Reviews Physics {\bf 1}, 264 (2019). 
\bibitem{11_wang} Fa Wang and T. Senthil, Phys. Rev. Lett. {\bf 106}, 136402 (2011).
\bibitem{13_watanabe} H. Watanabe, T. Shirakawa, and S. Yunoki, Phys. Rev. Lett. {\bf 110}, 027002 (2013).
\bibitem{10_chen} G. Chen, R. Pereira, and L. Balents, Phys. Rev. B {\bf 82}, 174440 (2010). 
\bibitem{11_chen} G. Chen and L. Balents, Phys. Rev. B {\bf 84}, 094420 (2011). 
\bibitem{11_dodds} T. Dodds, T.-P. Choy, and Y.B. Kim, Phys. Rev. B {\bf 84}, 104439 (2011). 
\bibitem{14_ishizuka} H. Ishizuka and L. Balents, Phys. Rev. B {\bf 90}, 184422 (2014). 
\bibitem{21_svoboda} C. Svoboda, W. Zhang, M. Randeria, and N. Trivedi, Phys. Rev. B {\bf 104}, 024437 (2021). 
\bibitem{17_romhanyi} J. Romhanyi, L. Balents, and G. Jackeli, Phys. Rev. Lett. {\bf 118}, 217202 (2017). 
\bibitem{20_paramekanti} A. Paramekanti, D.D. Maharaj, and B.D. Gaulin, Phys. Rev. B {\bf 101}, 054439 (2020). 
\bibitem{20_voleti} S. Voleti, D.D. Maharaj, B.D. Gaulin, G. Luke, and A. Paramekanti, Phys. Rev. B {\bf 101}, 155118 (2020). 
\bibitem{21_khaliullin} G. Khaliullin, D. Churchill, P.P. Stavropoulos, and H.-Y. Kee, Phys. Rev. Research {\bf 3}, 033163 (2021). 
\bibitem{21_mosca} D.F. Mosca, L.V. Pourovskii, B.H. Kim, P. Liu, S. Sanna, F. Boscherini, S. Khmelevskyi, and C. Franchini, Phys. Rev. B {\bf 103}, 104401 (2021). 
\bibitem{21_pourosvskii} L.V. Pourovskii, D.F. Mosca, and C. Franchini, Phys. Rev. Lett. {\bf 127}, 237201 (2021). 
\bibitem{10_aharen} T. Aharen, J.E. Greedan, C.A. Bridges, A.A. Aczel, J. Rodriguez, G.J. MacDougall, G.M. Luke, T. Imai, V.K. Michaelis, S. Kroeker, H.D. Zhou, C.R. Wiebe, and L.M.D. Cranswick, Phys. Rev. B {\bf 81}, 224409 (2010). 
\bibitem{10_devries} M.A. de Vries, A.C. Mclaughlin, and J.-W.G. Bos, Phys. Rev. Lett. {\bf 104}, 177202 (2010). 
\bibitem{20_hirai} D. Hirai, H. Sagayama, S. Gao, H. Ohsumi, G. Chen, T.H. Arima, and Z. Hiroi, Phys. Rev. Research {\bf 2}, 022063(R) (2020). 
\bibitem{20_maharaj} D.D. Maharaj, G. Sala, M.B. Stone, E. Kermarrec, C. Ritter, F. Fauth, C.A. Marjerrison, J.E. Greedan, A. Paramekanti, and B.D. Gaulin, Phys. Rev. Lett. {\bf 124}, 087206 (2020). 
\bibitem{21_lee} S. Lee, W. Lee, W. Guohua, J. Ma, H. Zhou, M. Lee, E.S. Choi, and K.-Y. Choi, Phys. Rev. B {\bf 103}, 224430 (2021). 
\bibitem{21_mustonen} O. Mustonen, H. Mutch, H.C. Walker, P.J. Baker, F.C. Coomer, R.S. Perry, C. Pughe, G.B.G. Stenning, C. Liu, S.E. Dutton, and E.J. Cussen, arXiv: 2108.02209 (unpublished). 
\bibitem{14_cao} G. Cao, T.F. Qi, L. Li, J. Terzic, S.J. Yuan, L.E. DeLong, G. Murthy, and R.K. Kaul, Phys. Rev. Lett. {\bf 112}, 056402 (2014). 
\bibitem{17_terzic} J. Terzic, H. Zheng, F. Ye, H.D. Zhao, P. Schlottmann, L.E. DeLong, S.J. Yuan, and G. Cao, Phys. Rev. B {\bf 96}, 064436 (2017). 
\bibitem{20_laguna} M.A. Laguna-Marco, E. Arias-Egido, C. Piquer, V. Cuartero, L. Hernandez-Lopez, P. Kayser, J.A. Alonso, J.A.T. Barker, G. Fabbris, C.A. Escanhoela Jr., and T. Irifune, Phys. Rev. B {\bf 101}, 014449 (2020). 
\bibitem{13_khaliullin} G. Khaliullin, Phys. Rev. Lett. {\bf 111}, 197201 (2013). 
\bibitem{15_bhowal} S. Bhowal, S. Baidya, I. Dasgupta, and T. Saha-Dasgupta, Phys. Rev. B {\bf 92}, 121113(R) (2015). 
\bibitem{15_meetei} O.N. Meetei, W.S. Cole, M. Randeria, and N. Trivedi, Phys. Rev. B {\bf 91}, 054412 (2015). 
\bibitem{17_jain} A. Jain, M. Krautloher, J. Porras, G.H. Ryu, D.P. Chen, D.L. Abernathy, J.T. Park, A. Ivanov, J. Chaloupka, G. Khaliullin, B. Keimer, and B.J. Kim, Nat. Phys. {\bf 13}, 633 (2017). 
\bibitem{19_kim} B.H. Kim, D.V. Efremov, and J. van den Brink, Phys. Rev. Mat. {\bf 3}, 014414 (2019). 
\bibitem{16_dey} T. Dey, A. Maljuk, D.V. Efremov, O. Kataeva, S. Gass, C.G.F. Blum, F. Steckel, D. Gruner, T. Ritschel, A.U.B. Wolter, J. Geck, C. Hess, K. Koepernik, J. van den Brink, S. Wurmehl, and B. Buchner, Phys. Rev. B {\bf 93}, 014434 (2016). 
\bibitem{16_pajskr} K. Pajskr, P. Novak, V. Pokorny, J. Kolorenc, R. Arita, J. Kunes, Phys. Rev. B {\bf 93}, 035129 (2016). 
\bibitem{16_phelan} B.F. Phelan, E.M. Seibel, D. Badoe Jr., W. Xie, and R.J. Cava, Solid State Communications {\bf 236}, 37 (2016). 
\bibitem{17_chen} Q. Chen, C. Svoboda, Q. Zheng, B.C. Sales, D.G. Mandrus, H.D. Zhou, J.-S. Zhou, D. McComb, M. Randeria, N. Trivedi, and J.-Q. Yan, Phys. Rev. B {\bf 96}, 144423 (2017). 
\bibitem{17_corredor} L.T. Corredor, G. Aslan-Cansever, M. Sturza, K. Manna, A. Maljuk, S. Gass, T. Dey, A.U.B. Wolter, O. Kataeva, A. Zimmermann, M. Geyer, C.G.F. Blum, S. Wurmehl, and B. Buchner, Phys. Rev. B {\bf 95}, 064418 (2017). 
\bibitem{17_hammerath} F. Hammerath, R. Sarkar, S. Kamusella, C. Baines, H.-H. Klauss, T. Dey, A. Maljuk, S. Ga$\beta$, A.U.B. Wolter, H.-J. Grafe, S. Wurmehl, and B. Buchner, Phys. Rev. B {\bf 96}, 165108 (2017). 
\bibitem{18_fuchs} S. Fuchs, T. Dey, G. Aslan-Cansever, A. Maljuk, S. Wurmehl, B. Buchner, and V. Kataev, Phys. Rev. Lett. {\bf 120}, 237204 (2018). 
\bibitem{18_gong} H. Gong, K. Kim, B.H. Kim, B. Kim, J. Kim, and B.I. Min, J. Mag. Magn. Materials {\bf 454}, 66 (2018). 
\bibitem{18_kusch} M. Kusch, V.M. Katkuri, N.A. Bogdanov, B. Buchner, T. Dey, D.V. Efremov, J.E. Hamann-Borrero, B.H. Kim, M. Krisch, A. Maljuk, M. Moretti Sala, S. Wurmehl, G. Aslan-Cansever, M. Sturza, L. Hozoi, J. van den Brink, and J. Geck, Phys. Rev. B {\bf 97}, 064421 (2018). 
\bibitem{18_nag} A. Nag, S. Bhowal, A. Chakraborty, M.M. Sala, A. Efimenko, F. Bert, P.K. Biswas, A.D. Hillier, M. Itoh, S.D. Kaushik, V. Siruguri, C. Meneghini, I. Dasgupta, and S. Ray, Phys. Rev. B {\bf 98}, 014431 (2018). 
\bibitem{17_yuan} B. Yuan, J.P. Clancy, A.M. Cook, C.M. Thompson, J. Greedan, G. Cao, B.C. Jeon, T.W. Noh, M.H. Upton, D. Casa, T. Gog, A. Paramekanti, and Y.-J. Kim, Phys. Rev. B {\bf 95}, 235114 (2017). 
\bibitem{18_calder} S. Calder, K. An, R. Boehler, C.R. dela Cruz, M.D. Frontzek, M. Guthrie, B. Haberl, A. Huq, S.A.J. Kimber, J. Liu, J.J. Molaison, J. Neuefeind, K. Page, A.M. dos Santos, K.M. Taddei, C. Tulk, and M.G. Tucker, Rev. Sci. Instrum. {\bf 89}, 092701 (2018). 
\bibitem{97_dereotier} P. Dalmas de Reotier and A. Yaouanc, J. Phys.: Condens. Matter {\bf 9}, 9113 (1997).
\bibitem{yaouanc_textbook} A. Yaouanc and P.Dalmas De Reotier, {\it  Muon Spin Rotation, Relaxation, and Resonance: Applications to Condensed Matter}, Vol. 147 (Oxford University Press, 2011).
\bibitem{15_ranjbar} B. Ranjbar, E. Reynolds, P. Kayser, B.J. Kennedy, J.R. Hester, and J.A. Kimpton, Inorg. Chem. {\bf 54}, 10468 (2015). 
\bibitem{99_wakeshima} M. Wakeshima, D. Harada, and Y. Hinatsu, Journal of Alloys and Compounds {\bf 287}, 130 (1999). 
\bibitem{17_kayser} P. Kayser, B.J. Kennedy, B. Ranjbar, J.A. Kimpton, and M. Avdeev, Inorg. Chem. {\bf 56}, 2204 (2017). 
\bibitem{93_rodriguez} J. Rodriguez-Carvajal, Physica B {\bf 192}, 55 (1993).
\bibitem{17_ziat} D. Ziat, A.A. Aczel, R. Sinclair, Q. Chen, H.D. Zhou, T.J. Williams, M.B. Stone, A. Verrier, and J.A. Quilliam, Phys. Rev. B {\bf 95}, 184424 (2017).
\bibitem{21_nguyen} L.T. Nguyen and R.J. Cava, Chem. Rev. {\bf 121}, 2935 (2021). 
\bibitem{16_nag} A. Nag, S. Middey, S. Bhowal, S.K. Panda, R. Mathieu, J.C. Orain, F. Bert, P. Mendels, P.G. Freeman, M. Mansson, H.M. Ronnow, M. Telling, P.K. Biswas, D. Sheptyakov, S.D. Kaushik, V. Siruguri, C. Meneghini, D.D. Sarma, I. Dasgupta, and S. Ray, Phys. Rev. Lett. {\bf 116}, 097205 (2016). 
\bibitem{18_nag_2} A. Nag, S. Bhowal, F. Bert, A.D. Hillier, M. Itoh, I. Carlomagno, C. Meneghini, T. Sarkar, R. Mathieu, I. Dasgupta, and S. Ray, Phys. Rev. B {\bf 97}, 064408 (2018). 
\bibitem{12_clancy} J.P. Clancy, N. Chen, C.Y. Kim, W.F. Chen, K.W. Plumb, B.C. Jeon, T.W. Noh, and Y.-J. Kim, Phys. Rev. B {\bf 86}, 195131 (2012).
\bibitem{19_aczel} A.A. Aczel, J.P. Clancy, Q. Chen, H.D. Zhou, D. Reig-i-Plessis, G.J. MacDougall, J.P.C. Ruff, M.H. Upton, Z. Islam, T.J. Williams, S. Calder, and J.-Q. Yan, Phys. Rev. B {\bf 99}, 134417 (2019). 
\bibitem{18_paramekanti} A. Paramekanti, D.J. Singh, B. Yuan, D. Casa, A. Said, Y.-J. Kim, and A.D. Christianson, Phys. Rev. B {\bf 97}, 235119 (2018). 
\bibitem{19_nag} A. Nag, S. Bhowal, M. Moretti Sala, A. Efimenko, I. Dasgupta, and S. Ray, Phys. Rev. Lett. {\bf 123}, 017201 (2019). 
\bibitem{uemura_book} S.L. Lee, R. Cywinski, and S.H. Kilcoyne, {\it Muon Science: Muons in Physics, Chemistry and Materials}, First Edition (CRC Press, 1999).
\bibitem{Slichter} C.P. Slichter, {\it Principles of Magnetic Resonance}, Third Enlarged and Updated Edition (Springer, 1989).

\end{thebibliography}
\end{document}